\documentclass[12pt]{article}
\pdfoutput=1
\usepackage{graphicx,amsmath,amssymb,amsfonts}
\usepackage{enumerate,setspace}
\usepackage{psfrag}
\usepackage{color}   
\usepackage{rotating}
\usepackage{pbox}
\usepackage[small, bf] {caption}
\usepackage{framed}
\usepackage{cite}
\makeatletter \renewcommand\@biblabel[1]{#1.~} \makeatother

\usepackage{makeidx}
\renewcommand{\a}{{\bf a}}
\usepackage{relsize}
\usepackage{bigints}
\usepackage{setspace}

\usepackage{hypernat}

\renewcommand{\a}{{\bf a}}

\newcommand{\eq}{{\rm eq}}

\newcommand{\st}{{\text{ST}}}

\renewcommand{\i}{k}

\usepackage{float}
\usepackage{fullpage}
\usepackage{lineno}

\newcommand{\vv}{\upsilon}
\renewcommand{\d}{\text{d}}

\newcommand{\EQ}{\begin{equation}}
\newcommand{\EE}{\end{equation}}
\newcommand{\EQA}{\begin{eqnarray}}
\newcommand{\EEA}{\end{eqnarray}}

\newcommand{\ad}{\text{ad}}
\newcommand{\stab}{\text{stab}}
\newcommand{\C}{{\cal C}}

\newcommand{\E}{\mathbf{E}}

\newcommand{\mf}{{\rm mf}}
\newcommand{\cE}{{\cal E}}
\renewcommand{\E}{{\bf E}}
\newcommand{\GGamma}{{\bf \Gamma}}

\begin{document}

\begin{titlepage}
\title{Pervasive adaptation of gene expression in Drosophila}
\author{Armita Nourmohammad$^{1,2*}$,  Joachim Rambeau$^{2,3}$,  Torsten Held$^{2}$, \\  Johannes Berg$^{2}$,  Michael L\"assig$^{2 *}$}

\date{\small $^1$ Joseph-Henri Laboratories of Physics and Lewis-Sigler Institute for Integrative Genomics, \\ Princeton University, Princeton, NJ 08544, USA\\ $^2$ Institut f\"ur Theoretische Physik, Universit\"at zu K\"oln, \\ Z\"ulpicherstr. 77, 50937, K\"oln, Germany \\ $^3$  Dynamique des Interactions Membranaires Normales et Pathologiques, UMR5235 CNRS,\\ Universit\'e de Montpellier, 34095 Montpellier, France}

\maketitle
\noindent

{\bf Gene expression levels are important molecular quantitative traits that link genotypes to molecular functions and fitness. 
In {\em Drosophila}, population-genetic studies in recent years have revealed substantial adaptive evolution at the genomic
 level. However, the evolutionary modes of gene expression
 have remained controversial. Here we present evidence that adaptation dominates the evolution of gene expression levels in flies. We show that {63\%} of the observed expression divergence across seven {\em Drosophila} species are adaptive changes driven by directional selection. Our results are derived from the variation of expression within species and the time-resolved divergence across a family of related species, using a new inference method for selection. Adaptive gene expression is stronger in specific functional classes, which include regulation,  sensory perception, sexual behavior, and morphology. Adaptation increases with broader codon usage and, consistently, highly expressed genes contribute less to adaptation. While most genes show coherent evolution in both sexes, we identify a large group of genes with sex-specific adaptation of expression, which predominantly occurs in males. Our analysis opens a new avenue to map system-wide selection on molecular quantitative traits independently of their genetic basis. }

\vspace{3.5 cm}
\vfill{}
\noindent \footnotesize{$^*$ Correspondence should be addressed to: Armita Nourmohammad (armitan@princeton.edu) or Michael L\"assig (mlaessig@uni-koeln.de).}

\end{titlepage}

In recent years, several studies have found evidence for widespread adaptive evolution of the {\em Drosophila} genome~\cite{Andolfatto:2005ir,Lassig:2007vb,Sella:2009hs}. This includes adaptive changes in non-coding sequence, consistent with classical ideas on the importance of regulatory evolution for phenotypic adaptation~\cite{King:1975gx}.  Gene expression levels are important molecular phenotypes that quantify the effects of regulation on organismic traits and fitness. Insights on how genome evolution affects gene expression have come from studies of quantitative trait loci (QTL); see refs.~\cite{Fraser:2011kt, Romero:2012hu,Pai:2015pg} for recent reviews. In yeast, at least 10\% of the genes have been inferred to undergo adaptive evolution of expression~\cite{Fraser:2010gm}. In flies, expression-QTL analysis has been used to estimate cis- and trans- effects on expression~\cite{Genissel:2007dr,Wittkopp:2008ki} and to compare the evolution of expression and that of the underlying regulatory sequence~\cite{Coolon:2014hk}; related studies have been performed in yeast~\cite{Fraser:2014af}.  However, given the complexity of the regulatory genotype-to-phenotype map and the limited sensitivity of QTL studies, our understanding of how adaptive genome changes relate to mRNA and protein levels has remained incomplete~\cite{Hoekstra:2007fi,Fraser:2011kt,Pai:2015pg}. 

An alternative approach is to analyze the evolution of gene expression by methods of quantitative genetics, without explicit reference to genetic evolution of the QTL~\cite{Rifkin:2003cj, Lemos:2005wz, Khaitovich:2005gr,Gilad:2006kta,  Whitehead:2006dj, Zhang:2007bf, Bedford:2009fy, Fraser:2011kn,Romero:2012hu,Pai:2015pg}. 
These studies compare the expression divergence across species, the variation within species, and the expected behavior for neutral evolution~\cite{Lynch:1986tu}. 
 A broad picture of evolutionary constraint on gene expression levels caused by stabilizing selection has  emerged in a number of species, including {\em Drosophila}~\cite{Rifkin:2003cj, Lemos:2005wz, Gilad:2006kta,Bedford:2009fy,Romero:2012hu}.  A comparative study between human and chimpanzee has produced signatures of predominantly neutral evolution of gene expression~\cite{Khaitovich:2005gr}. Other studies in primates have identified stabilizing selection, as well as lineage- and tissue-specific directional expression changes~\cite{Gilad:2006kta, Zhang:2007bf, Blekhman:2008er,Fraser:2011kn, Brawand:2011,Romero:2012hu,Fraser:2013}. However, it has remained difficult to demonstrate that positive selection, as opposed to relaxed stabilizing selection, is the evolutionary cause of expression divergence~\cite{Fraser:2011kt}. Thus, estimating the genome-wide contribution of adaptation to the evolution of gene expression is an outstanding problem.

In this paper, we show that adaptation is the prevalent evolutionary mode of gene expression throughout the {\em Drosophila} genus. Our method is based on recent theoretical results on the evolution of molecular quantitative traits~\cite{Held:2014ve, Nourmohammad:2013ty, NourMohammad:2013in} and has two essential features. First, we infer adaptation driven by directional selection together with conservation under stabilizing selection, which allows us to discriminate these two modes in the evolution of gene expression. Second, we identify observables that decouple from number and effects of the underlying QTL. These molecular determinants of gene expression are often unknown, vary considerably between genes, and confound a naive phenotype-based inference of selection.

\subsection*{Results} 

\subsubsection*{The pattern of gene expression divergence}

We use gene expression data from samples of inbred male and female individuals~\cite{Zhang:2007bf}, which cover 6332 orthologous genes in seven {\em Drosophila} species. A phylogenetic tree of these species is shown in  Fig.~\ref{clade_phyl}.  Gene expression levels are defined in a standard way as logarithms of mRNA counts, suitably normalized to account for differences in assay sensitivity between experimental probes (Materials and Methods). Within each population, we use these data to estimate the mean expression level of a gene, its total heritable expression variance (referred to as expression diversity), and its non-heritable expression variance between inbred individuals. For each pair of populations, we obtain the cross-species expression divergence of a gene as the squared difference between the population mean levels (see Materials and Methods for details). These quantities define the (suitably scaled) divergence-diversity ratio $\Omega$, which plays a key role in our evolutionary analysis (Box~1 and Materials and Methods). For a given gene, the evolution of the $\Omega$ ratio depends only weakly on the effect distribution of its expression QTL and on the amount of recombination between these loci. For neutral evolution, this property is implicitly contained in classical quantitative genetics approaches~\cite{Lynch:1986tu, Leinonen:2013ht}, but it holds more generally under stabilizing and directional selection~\cite{Nourmohammad:2013ty,NourMohammad:2013in,Held:2014ve}.

To obtain a genome-wide evolutionary picture of gene expression in {\em Drosophila}, we evaluate the aggregate divergence-diversity ratio for all genes in our data set (Materials and Methods). Grouping the species into 6 clades, we obtain a consistent pattern of the $\Omega$ ratio as a function of divergence time, $\tau$ (Fig.~1A). These data show macro-evolution of expression levels: already the average expression divergence for the youngest clade, {\em D.~melanogaster} and {\em D.~simulans}, is by a factor 2 larger than the average diversity within species (Fig.~\ref{clade_phyl}). On the other hand, the observed $\Omega$ values for all clades are substantially smaller than the expected values under neutral evolution, which can be computed analytically~\cite{Held:2014ve}. This constraint suggests that gene expression evolves under stabilizing selection, in agreement with previous studies~\cite{Rifkin:2003cj,Lemos:2005wz,Bedford:2009fy}  and with a standard $Q_\st / F_\st$ analysis~\cite{Leinonen:2013ht} (Materials and Methods). Importantly, however, the $\Omega$ ratio increases with divergence time throughout the {\em Drosophila} genus and does not show evidence of saturation for larger values of $\tau$,  in accordance with a similar pattern of the expression divergence observed previously\cite{Zhang:2007bf}. In the following, we will show that this feature reflects adaptive evolution of gene expression.

\subsubsection*{Fitness model for gene expression}
The inference of adaptation is based on a minimal dynamical model of selection: gene expression levels evolve in a single-peak fitness seascape~\cite{NourMohammad:2013in,Held:2014ve}. This model is illustrated in Box~1 and formally defined in Materials and Methods. The fitness peak for a given gene performs a random walk over macro-evolutionary periods. This walk maps continual changes of the optimal expression of that gene, which are generated by long-term environmental shifts and epistatic co-evolution with other genes. Despite its simplicity, the seascape model combines two salient features of selection on gene expression: stabilizing selection generates evolutionary constraint, and directional selection drives long-term adaptive changes. These selection components are measured by two parameters, the stabilizing strength $c$ and the driving rate~$\vv$. 

Here we use the dependence of the divergence-diversity ratio on evolutionary time, $\Omega (\tau)$, to infer the fitness seascape of gene expression (Box~1). This method discriminates directional selection in a genuine fitness seascape from purely stabilizing selection in a static fitness landscape, providing a more powerful inference of adaptive evolution than $Q_\st / F_\st$ analysis~\cite{Leinonen:2013ht} (Materials and Methods). It also estimates the most important summary statistics of the adaptive process: the cumulative fitness flux $\Phi$, which measures the fitness gain through adaptive expression changes over an evolutionary period (Box~1, Materials and Methods).

\subsubsection*{The fitness seascape of {\em Drosophila} gene expression} 
We first use the aggregate divergence-diversity data to infer a gene-averaged fitness seascape of expression levels in {\em Drosophila} (Fig.~1A). The least-square fitted seascape model (green line) contains stabilizing and directional selection, leading to adaptive evolution. This model explains the observed pattern $\Omega (\tau)$: the short-term evolutionary constraint is caused by stabilizing selection, and the approximately linear long-term increase signals adaptation. Because genetic drift and adaptation differ in tempo, the contribution of adaptation to expression divergence depends on evolutionary time: the adaptive part is small for the youngest species clades, but adaptation becomes dominant across the entire {\em Drosophila} genus (green shaded area). In contrast, stabilizing selection alone cannot explain the {\em Drosophila} expression data. In a static fitness landscape, genetic drift generates a rapidly saturating pattern of $\Omega (\tau)$ that is not observed in the data (Fig.~\ref{fig:Hartl}).

We can extend the seascape inference to individual genes, using a Bayesian inference scheme that decouples from number and effects of their expression QTL (Materials and Methods). We obtain a posterior distribution of stabilizing strength and fitness flux for each gene. For 54\% of all genes, we infer a significant cumulative fitness flux $\Phi$ across the {\em Drosophila} genus; we classify these genes as adaptively regulated (Table~S1, Materials and Methods). Fig.~1B shows the average posterior distribution of the fitness flux, which determines a clade-specific adaptive fraction of the expression divergence (Table~1, Materials and Methods). This fraction increases with clade divergence time, in accordance with the aggregate data of Fig.~1A. Between {\em D.~melanogaster} and {\em D.~simulans}, which diverged about 2-3 Myrs ago, {92\%} of the expression divergence can be attributed to genetic drift under stabilizing selection. Across the entire {\em Drosophila} genus, which has its last common ancestor about 40 Myrs ago, we infer {63\%} of the expression divergence to be adaptive. The Bayesian scheme also allows us to quantify the overall statistical significance of our selection inference. In Fig.~1C, we plot the cumulative log-likelihood score  for all genes as a function of  $c$ and $\Phi$. As shown by a log-likelihood test, the global maximum-likelihood seascape model is strongly favored over the maximum-likelihood landscape model ($P < 10^{-3600}$) and over neutral evolution ($P < 10^{-5400}$)  (Materials and Methods). We note that this analysis rejects neutral evolution and evolution under static stabilizing selection independently of model assumptions on the adaptive dynamics. 

\subsubsection*{Testing alternative evolutionary scenarios}
The minimal seascape model explains the pattern of gene expression divergence across the {\em Drosophila} genus in a parsimonious way. But are there equally parsimonious alternative modes of selection or demography that are consistent with the data? To assess the specificity and robustness of the seascape-based inference, we characterize the statistics of gene expression levels in a number of alternative modes of evolution by analytical approximations and simulations, and we compare the results to the  {\em Drosophila} data.
 
First, demographic effects may increase or decrease the effective population size in a specific lineage, which affects the stabilizing strength $c$ for all genes. As shown in Fig.~\ref{fig:demographic}, lineage-specific changes in effective population size that persist over sufficiently long evolutionary periods can be traced in the aggregate divergence-diversity function $\Omega(\tau)$. Such effects are not observed in our data, which suggests that long-term demographic effects do not play a dominant role in the evolution of {\em Drosophila} gene expression levels (Fig.~\ref{fig:demographic}). This result does not exclude short-term changes of population size, which occurred, for example, in the recent evolution of the {\em D. melanogaster} lineage~\cite{Lachaise:1988tg}. Such changes can be traced in sequence polymorphism spectra~\cite{Glinka:2003tr,Haddrill:2005,Stephan:2007gc,Thornton:2007bl}, but they have only minor effects on gene expression levels.

Next, we ask if the {\em Drosophila} data can be explained by lineage- and gene-specific relaxation of stabilizing selection. We consider a specific non-adaptive mode of expression changes: functional genes evolve under stabilizing selection in a static fitness landscape, but individual genes can (partially) lose function at a given point in their evolutionary history, which relaxes selection on their expression. We model loss of function as stochastic events occurring at a small rate, independently for each gene and on each lineage.  This model produces a divergence-diversity function $\Omega (\tau)$ with a long-term nonlinearity that is not seen in the $\Omega$ data (Fig.~\ref{fig:Alternative2}). The most direct way to discriminate between relaxation of selection and adaptive evolution is to use a directional bias: most functional genes are up-regulated by stabilizing selection (a similar bias has been exploited in expression QTL studies~\cite{Fraser:2010gm, Fraser:2011kt,Fraser:2013}). In the loss-of-function mode, a comparison of expression levels for a given gene should show small cross-species differences at higher expression levels (i.e., between the lineages with a functional gene) together with large deviations at lower levels (i.e, in the lineages with lost gene function). Accordingly, the distribution of expression divergence values for a given species pair should show a broad tail generated by the loss events. These features are not observed in our data, indicating that relaxed stabilizing selection alone cannot explain the evolution of {\em Drosophila} expression levels (Fig.~\ref{fig:Alternative2}).  Of course, loss of gene function does happen in our phylogeny, but the affected genes will often lose expression altogether and, hence, will be suppressed in our data set. 

We can also compare the {\em Drosophila} data with alternative models of adaptive evolution. For example, individual genes can undergo a (partial) neo-functionalization that requires a major change in their expression. We describe this mode of evolution by a punctuated fitness seascape, in which large shifts of the peak position are stochastic events occurring at a small rate~\cite{Held:2014ve}. This process produces an aggregate divergence-diversity function $\Omega (\tau)$ that is compatible with the data, but a broad tail in the distribution of expression divergence values that is not observed (Fig.~\ref{fig:Alternative2}). We conclude that gradual but continual changes in optimal levels, as described by our minimal model, are the dominant evolutionary force driving the adaptation of gene expression in {\em Drosophila}. 

\subsubsection*{Functional and mechanistic determinants of selection}
By applying our inference to specific classes of genes, we can get a more detailed view on adaptation of gene expression in {\em Drosophila.} First, we observe a strong correlation between codon usage and adaptation: genes with specific codons show strongly reduced adaptive expression divergence and lower average fitness flux than genes with broad codon usage (Fig.~2A,B, Table~\ref{table:fraction}). Specific codon usage is known to be prevalent in highly expressed genes~\cite{Ikemura:1985ui}; consistently, we find stronger conservation of expression and lower levels of fitness flux in this class (Table~\ref{table:fraction}). Different codons for the same amino acid differ in their efficiency of translation~\cite{Ikemura:1985ui,SHIELDS:1988uj}, which implies that genes with broad codon usage have a higher potential for adaptive expression changes at the post-transcriptional level. Here we find stronger adaptation at the mRNA level in this gene class, which suggests a two-tier mode of evolution: adaptive mRNA changes lay the ground on which coherent adaptive tuning of protein levels can build. 

At the same time, we find no significant correlation between the fitness flux for expression changes and adaptive evolution of amino acid sequence, as measured by a McDonald-Kreitman test~\cite{Kreitman:1991vh} (Fig.~\ref{fig:MK}). We conclude that gene expression and gene function provide two largely independent modes of evolution. For a metabolite or a transcription factor, adaptive changes of its cellular concentration are often coupled with conservation of its function. 

Our gene-specific inference can be used to detect functional gene classes associated with adaptive evolution of regulation.  A full ranking of gene classes by enrichment in adaptively regulated genes with associated $P$ values is reported in {Table~S1}. Gene functions associated with enhanced adaptive evolution of expression include  sensory perception, regulation, neural maturation,  regulation of growth, aging and morphology. Adaptively regulated functions also include response to UV radiation, which has recently been identified as an important climate-mediated trait in humans~\cite{Hancock:2011,Fraser:2013}. Adaptive evolution of genes related to growth, regulation and morphology has been previously inferred by expression QTL and comparative studies of gene regulation in other species~\cite{Fraser:2011kn,Fraser:2011kt,Romero:2012hu}. Here we identify these categories from a quantitative, system-wide scan for adaptively regulated genes. This points to the power of our phenotype-based inference scheme, which is not confounded by the combinatorial complexity of cis-regulatory sequence in higher eukaryotes.

\subsubsection*{Sex-specific evolution of expression}
We can also test the role of expression differentiation between male and female individuals for adaptive evolution across the {\em Drosophila} genus. The sex specificity of a given gene~\cite{Zhang:2007bf}, defined as the difference between its male and female expression level $E_{\mf} =E_m-E_f$, is a distinct quantitative trait whose evolutionary pattern can be analyzed by our method. We can distinguish two modes of evolution: conservation of  sex specificity maintained by stabilizing selection and sex-specific adaptation of expression (Fig.~2C). Most genes of our data set have well-conserved and often small sex specificity; these genes evolve their expression levels coherently between males and females~\cite{Zhang:2007bf}. The remaining 19\% of the genes have a significant cumulative fitness flux of their specificity trait, $\Phi_{\mf}$; we classify them as undergoing sex-specific adaptation of expression in the {\em Drosophila} genus. These genes cover all four chromosomes of the {\em Drosophila} genome. 

Gene functions associated with  sex-specific adaptation of expression include regulation of translation, reproduction, post-mating behavior and (immune) response to biotic stimuli (Table~S2). To understand the distribution of these adaptive processes between sexes, we apply our inference to classes of genes with different species-averaged sex bias of expression{~\cite{Assis:2012ez}}. For male-biased genes, the divergence-diversity ratio $\Omega_{\mf}$ signals substantial sex-specific adaptation (Fig.~2D). Consistently, the fitness flux  $\Phi_{\mf}$ is strongly enhanced in genes that are predominantly expressed in males (Fig.~2E). The fitness flux is lower in the other classes, including genes expressed predominantly in females. Together, we find a remarkable evolutionary asymmetry between sexes: male bias in expression is associated with adaptive evolution of expression (orange shaded areas in Fig.~2D,E); female bias in expression is under weaker directional selection, which suggests it primarily reflects conserved physiological differences between male and female organisms. This result complements a previously observed evolutionary asymmetry at the sequence level:  genes with male-biased expression show increased amino acid divergence~\cite{Zhang:2007bf}. As suggested by a McDonald-Kreitman test, this increase can be associated with adaptive evolution of gene function (Fig.~\ref{fig:MK}).

\subsection*{Discussion} 
We have shown that adaptive regulation accounts for most of the macro-evolutionary divergence in gene expression across the {\em Drosophila} genus. Genes differ considerably in the amount of adaptation, depending on their codon usage, sexual differentiation, and functional class. These results provide evidence for system-wide adaptation of gene regulation in {\em Drosophila} already at the primary level of transcription, notwithstanding further evolutionary complexities at the level of translation~\cite{Romero:2012hu,Fraser:2014af}. It remains to be seen whether a similar prevalence of adaptation in the evolution of expression will be found in different species.

Our inference of adaptation is based on the expression divergence-diversity ratio, which depends on the evolutionary distance between species. It exploits two fundamental evolutionary features of quantitative traits: at short evolutionary distance, the divergence is always near neutrality; at longer distance, it depends jointly on stabilizing and directional selection. These features generate a divergence pattern with two distinct molecular clocks, as shown in Fig.~1. Importantly, the phenotypic evolution of gene expression decouples from details of its genetic basis. This explains why we find strong selection on gene expression levels although selection on individual QTL is often weak~\cite{Sunyaev:2013}. 

Our method can be applied to a broad spectrum of molecular quantitative traits with a complex genetic basis, provided comparative data from multiple, sufficiently diverged species are available. Such traits include genome-wide protein levels, protein-DNA binding interactions or enzymatic activities. For most of these traits, we have only partial knowledge of the underlying genetic loci and their effects on trait and fitness. Our method complements QTL studies and opens a way to infer quantitative phenotype-fitness maps at the systems level.

\singlespacing
\begin{center}
\framebox{
 \parbox{0.69 \textwidth}{
\includegraphics[height=3.8cm]{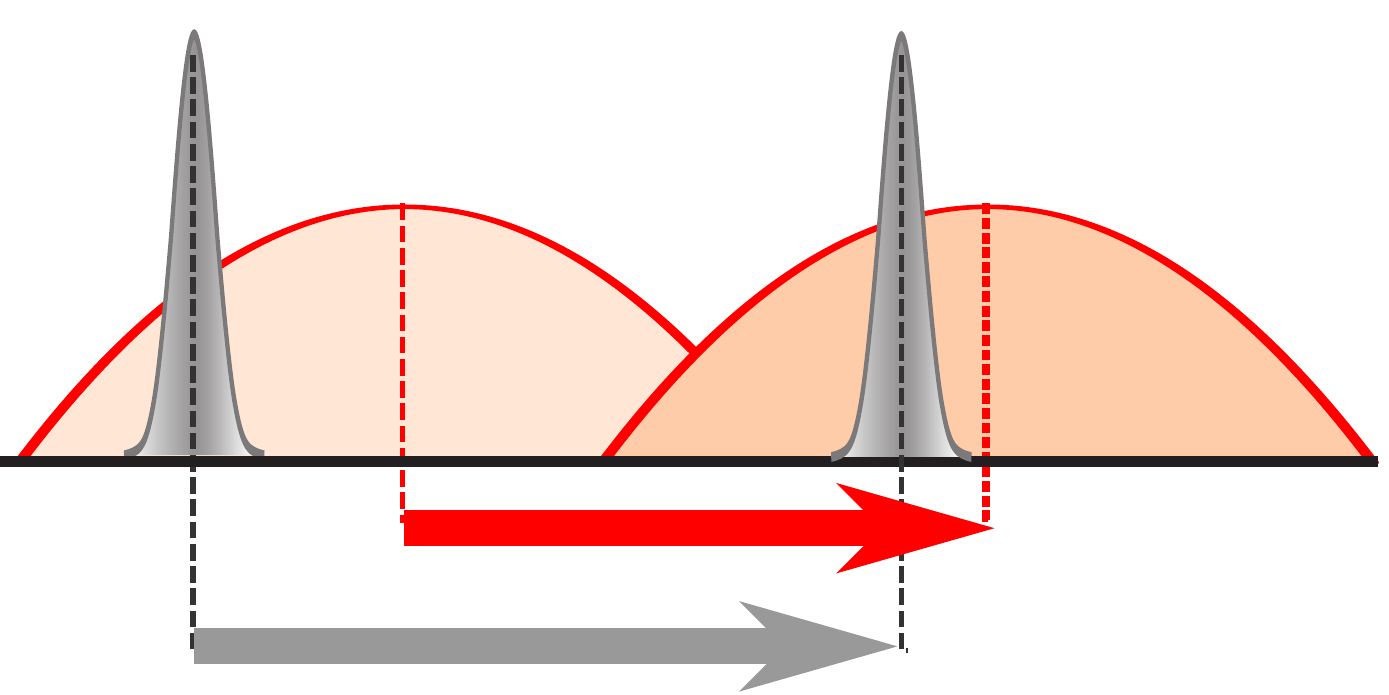}

\small {\small \bf Box 1: Inferring adaptive evolution of quantitative traits.} 
The schematic shows the evolution of a quantitative trait in a single-peak fitness seascape. The distribution of trait values within a species (grey curves) changes over macro-evolutionary periods, which can be observed as cross-species divergence of the mean trait values (grey arrow). In the underlying fitness seascape (red curves), evolutionary displacements of the fitness peak lead to lineage-specific optimal trait values and directional selection (red arrow). The minimal fitness model has two parameters: the stabilizing strength $c$ is  proportional to the inverse square width of the fitness peak, the driving rate $\vv$ measures the mean square displacement of the fitness peak per unit of evolutionary time (see Materials and Methods for precise definitions). 
We infer selection on quantitative traits from their time-resolved inter-species divergence, $D(\tau)$, and their intra-population diversity, $\Delta$, as defined in Materials and Methods. We evaluate the ratio $\Omega (\tau) =  \pi_0 D(\tau) / \Delta$, scaled by the neutral sequence diversity $\pi_0$ and averaged over a family of traits for three or more species with different divergence times $\tau$ (Fig.~1A, squares). The $\Omega$ test~\cite{Held:2014ve} is guided by theoretical results on the evolution of quantitative traits~\cite{ NourMohammad:2013in, Held:2014ve}. These provide the analytical form of  $\Omega (\tau)$ in a fitness seascape ($c > 0$, $\vv > 0$; green solid line). The corresponding form $\Omega_\eq (\tau)$ in a fitness landscape of the same stabilizing strength ($c > 0$, $\vv = 0$; blue solid line) reaches a saturation value $\Omega_\stab$. Fitting the seascape model to the $\Omega$ data determines the decomposition $\Omega (\tau) = \Omega_\eq (\tau) + \Omega_\ad (\tau)$ (blue and green shaded areas). The amplitude ratio $\omega_\ad (\tau) = \Omega_\ad (\tau) / \Omega (\tau)$ gives the fraction of trait divergence that is adaptive, i.e., driven by directional selection. In the linear regime $\Omega (\tau) \approx \Omega_\stab + \Omega_\ad (\tau)$, which covers all species clades in this dataset, the fitted amplitudes provide simple estimates of the selection parameters, $c \approx 1/(2 \Omega_\stab)$ and $\vv  \approx 2\Omega_\ad (\tau) / \tau$, and of the resulting cumulative fitness flux $2N \Phi =2 c \vv \tau$ (scaled by the effective population size $N$). The divergence-diversity ratio for neutral evolution, $\Omega_0 (\tau)$, is shown for reference ($c = 0$; grey solid line). A Bayesian extension of the $\Omega$ test and its  relationship to other trait-  and sequence-based selection tests ($Q_{\rm st} /F_{\rm st}$ test~\cite{Leinonen:2013ht}
, Ornstein-Uhlenbeck models,~\cite{Hansen:1997ws,Bedford:2009fy,Kalinka:2010dw,Rohlfs:2014bl} and 
McDonald-Kreitman test~\cite{Kreitman:1991vh})
 are discussed in Materials and Methods.}  
} 
\end{center}

\newpage
\begin{figure}[h!]
\begin{center}
\includegraphics[width=0.7 \textwidth]{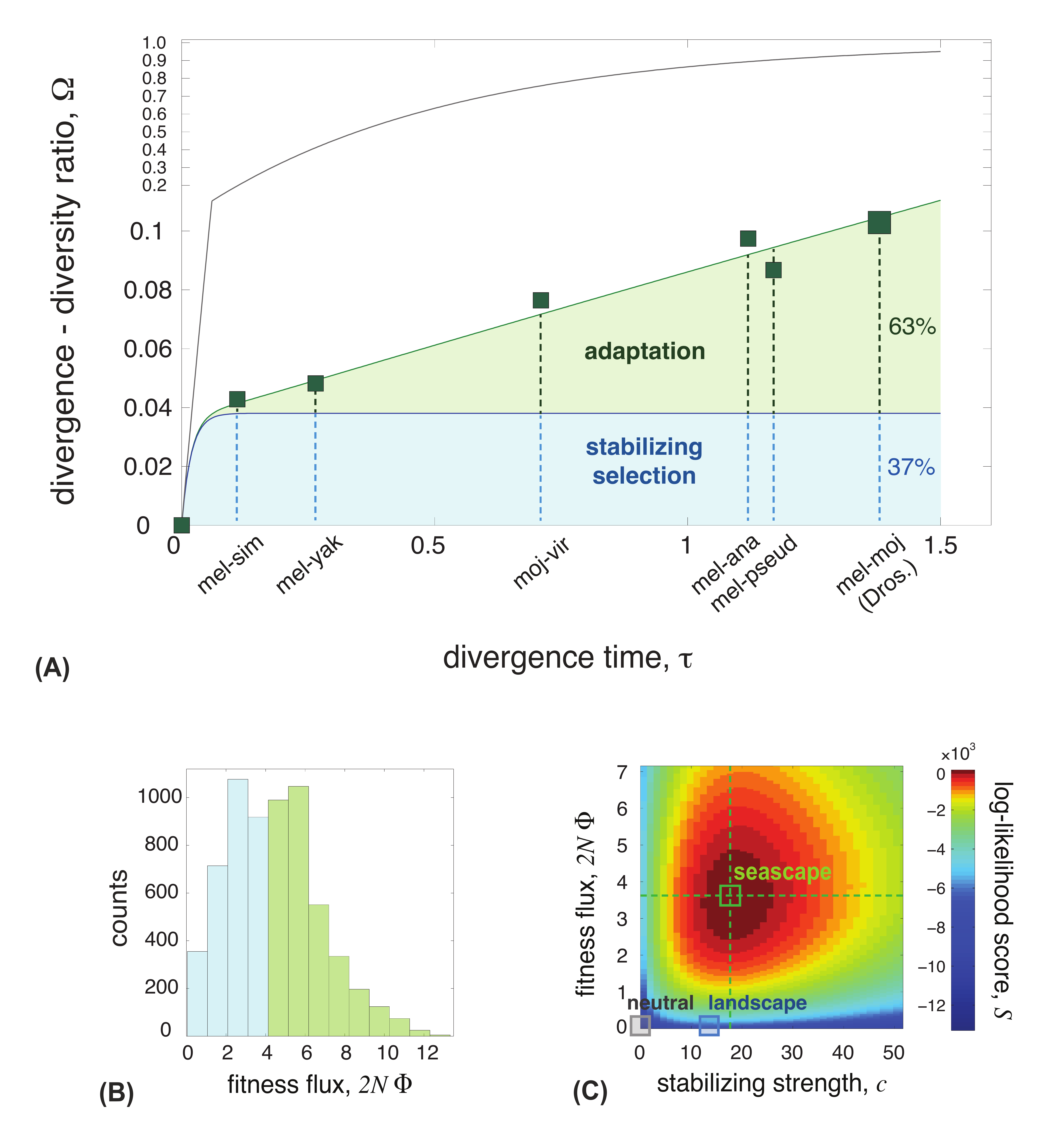}
\end{center}
\caption{{\bf Adaptive evolution of gene expression.}
{\bf (A)}~ The aggregate divergence-diversity ratio, $\Omega$, from all genes of this data set is plotted against the divergence time, $\tau$ for six partial species clades ({\scriptsize$\blacksquare$}) and for the entire {\em Drosophila} genus (${\scriptsize \blacksquare}$). Species clades and divergence times (scaled by the rate of synonymous mutations) are defined by the phylogenetic tree of the {\em Drosophila} Reference Panel (Fig.~\ref{clade_phyl}A and Materials and Methods). These data are shown together with theoretical curves $\Omega (\tau)$ under directional selection (green line), under stabilizing selection (blue line), and for neutral evolution (grey line). Inferred model parameters are $c^* = 18.6$ and $\vv^* = 0.07$; see Box~1 and Materials and Methods for model details. We infer a time-dependent adaptive component of the expression divergence (green shaded area); the complementary component is generated by genetic drift under stabilizing selection (blue shaded area).  Adaptation accounts for {63\%} of the expression divergence across the {\em Drosophila} genus. See Fig.~\ref{fig:Hartl} for a comparison of the same data to models of time-independent stabilizing selection~\cite{Bedford:2009fy}. 
{\bf (B)}~Distribution of maximum-likelihood values of the scaled cumulative fitness flux, $2N \Phi$, inferred for individual genes (Materials and Methods).  Our inference classifies $54\%$ of all genes as adaptively regulated ($2N \Phi > 4$, green shaded part of the distribution). 
{\bf (C)}~Bayesian inference of fitness models. The posterior log-likelihood score $S(c, \Phi)$ favors the optimal seascape model $(c^* = 18.6, 2N\Phi^*= 3.6$; green square) over the best landscape model ($c_\eq^* = 16, \Phi =0$; blue square) and the neutral model ($c = 0, \Phi = 0$; grey square).
}
\label{Fig:OmegaCurves} 
\end{figure}

\newpage 
\begin{figure}[h!]
\includegraphics[width=\textwidth]{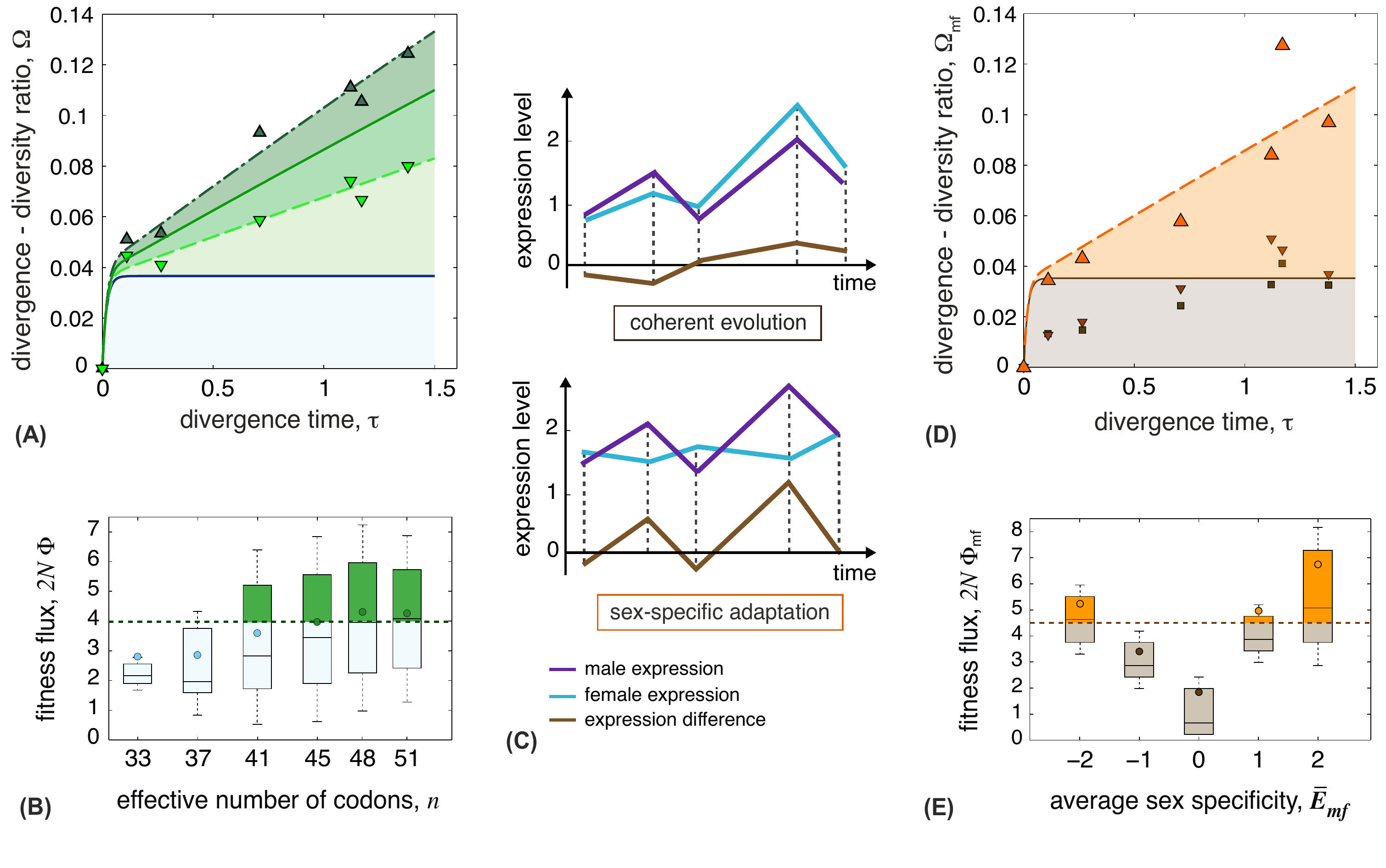}
\caption{
{\bf Adaptation depends on codon usage and sex bias.}
{\bf (A)} The aggregate divergence-diversity ratio, $\Omega$, for genes with broad codon usage ($\triangle$) and for genes with specific codon usage ($\bigtriangledown$) is shown together with theoretical curves under directional selection (dashed and dashed-dotted lines); the theoretical curve inferred for all genes is shown for comparison (solid line, cf.~Fig.~1A). Codon usage is measured by the effective number of codons~\cite{Wright:1990tf} (Materials and Methods); inferred model parameters are listed in Table~\ref{table:fraction}.
{\bf (B)}~The distribution of the cumulative fitness flux, $ 2N \Phi$, is plotted against the effective number of codons~\cite{Wright:1990tf}, $n$ (Materials and Methods; circle: average, line: median, box: 50\% around median, bars: 70\% around median).
{\bf (C)} Conservation of sex specificity (top panel) vs.~sex-specific adaptation of expression (bottom panel). The sex-specificity trait (brown line) is defined as the difference between male  and female expression levels (purple and blue lines). The schematics show all three lines as functions of evolutionary time. 
{\bf (D)} The aggregate divergence-diversity ratio of the sex-specificity trait, $\Omega_{\rm mf}$, for all genes ($\square$), genes with male-biased expression ($\triangle$), and genes with female-biased expression ($\bigtriangledown$), is shown together with theoretical curves under directional selection. 
{\bf (E)} The distribution of the cumulative fitness flux for the sex-specificity trait,  $ 2N \Phi_\mf$, is plotted against the species-averaged sex specificity, $\bar E_\mf$ (Materials and Methods). Sex-specific adaptation ($2N \Phi_\mf>4.5$, orange shaded part) occurs predominantly in male-biased genes. 
}
\end{figure}

\newpage
\subsection*{Materials and Methods}

\subsubsection*{Summary}
{\small
Synonymous genome sequence is used to estimate the neutral sequence diversity $\pi_0 = 4\mu N$ (ref.~\cite{Mackay:2012fd}, $\mu$ is the point mutation rate and $N$ the effective population size) and the species divergence times $\tau_{ij}$ (ref.~\cite{DrosophilaGenomesConsortium:2007gf}, scaled in units of $\mu^{-1}$); these data underlie the phylogenetic tree~\cite{DrosophilaGenomesConsortium:2007gf} (Fig.~\ref{clade_phyl}). The  expression levels $E^\alpha_{i,s,\kappa}$ measured by Zhang et al.~\cite{Zhang:2007bf} are labelled by gene $\alpha$, species $i$, sex $s$, and replicates $\kappa$ of an inbred  line. The levels are normalized to mean 0 and cross-gene variance 1 for each individual; the effects of this normalization on our analysis is tested in Fig.~\ref{fig:rescaling}. For a given gene $\alpha$, we estimate isogenic variance between experimental replicates of an inbred line $\delta^\alpha_i$ and genetic mean $\Gamma^\alpha_i$  in each species, and the divergence $D^\alpha_{ij} = (\Gamma^\alpha_i - \Gamma^\alpha_j)^2$ between any two species $i,j$; we also estimate genetic variance (expression diversity) $\Delta^\alpha$ using data from the two distinct strains of  {\em D.~simulans}. We define the aggregate divergence-diversity ratio $\Omega^\alpha_{ij} =  \pi_0 \langle D_{ij} \rangle / \langle \Delta \rangle$, where angular brackets denote averages over genes. For each clade $\C$ in our phylogeny, we obtain the aggregate data $(\tau_\C, \Omega_\C)$ shown in Figs.~1A, 2A by averaging $\tau_{ij}$ and $\Omega_{ij}$ over all pairs of species $i,j \in \C$ that are connected via the root of the clade.

The minimal fitness seascape for a given gene takes the form 
\[
f(E,t) = - \frac{c}{2N E_0^2} \big (E - E^*(t) \big)^2,
\]
where the optimal trait value $E^*(t)$ performs an Ornstein-Uhlenbeck process with mean square displacement $\vv E_0^2$ per unit $\mu^{-1}$ of evolutionary time; $E_0^2$ is the average genetic variation of expression in the long-term limit of neutral evolution~\cite{Nourmohammad:2013ty}. This non-equilibrium model generates adaptive evolution with an average scaled cumulative fitness flux $\langle 2N \Phi \rangle = 2 c \vv \, \tau_{\, \rm Dros.}$ across the {\em Drosophila} genus ($\tau_{\, \rm Dros.} = 1.4$; Fig.~\ref{clade_phyl}). Applying the $\Omega$ test (Box~1) to aggregate expression data $(\tau_\C, \Omega_\C)$, we infer a global fitness seascape with parameters ($c^* = 18.6, \vv^* = 0.07$) and an average fitness flux $2N \Phi^* = 3.6$ per gene. Control fits of the same data to equilibrium models, including the well-known Ornstein-Uhlenbeck  dynamics for the population mean trait~\cite{Hansen:1997ws, Bedford:2009fy, Kalinka:2010dw,Rohlfs:2014bl}, are shown in Fig.~\ref{fig:Hartl}. The probabilistic extension of this test evaluates the Bayesian posterior probability distribution $Q(c,  \Phi | \E^\alpha)$ of individual genes, given their sample mean data $\E^\alpha  = (E^\alpha_1, \dots, E^\alpha_7)$. This produces gene-specific expectation values $c^\alpha$ and $\Phi^\alpha$ (Figs.~1B and 2B); we use the condition $2N \Phi^\alpha > 4$ to infer adaptively regulated genes (Table~S1). The cumulative log-likelihood score $S(c, \Phi) = \sum_\alpha \log Q(c, \Phi | \E^\alpha)$ serves to quantify the statistical significance of our inference (Fig.~1C). 

To test for lineage-specific demographic effects, we compare the aggregate $\Omega$ data to theoretical functions $\Omega (\tau, \tau_i)$ computed for an alternative model with a change in effective population size on the phylogenetic branch of species $i$ (Fig.~\ref{fig:demographic}). We also examine two alternative selection scenarios: relaxed stabilizing selection by partial loss of function ($c^\alpha$ switches to a reduced value with rate $\gamma$) and punctuated fitness peak shifts ($E^{* \alpha}$ jumps by an amount of order $E_0$ with a rate of order $\vv \mu$) (Fig.~\ref{fig:Alternative2}). The observed distributions of cross-species expression differences are consistent with the minimal seascape model but at variance with both alternative models (Fig.~\ref{fig:Alternative2}). 

To infer sex-specific evolution, we define specificity traits as differences between male and female expression levels, $E^\alpha_{\mf, i} = E^\alpha_{m,i} - E^\alpha_{f,i}$, for each gene~\cite{Zhang:2007bf}. Genes with sex-specific adaptive evolution of expression are identified by a condition on the cumulative fitness flux for the specificity trait, $2N\Phi_\mf^\alpha > 4.5$ (Table~S2). Genes with male- and female-biased expression are identified using the results of ref.~\cite{Assis:2012ez}. 

We simulate  Fisher-Wright evolution in fitness land- and seascapes to validate our probabilistic inference scheme and to establish its robustness under trait epistasis (Fig.~\ref{fig:sim}). 
}

\subsubsection*{1. Data and primary analysis} 

\paragraph{Sequence data and phylogeny.}
Our inference procedure requires the following global sequence-based information (which does not include expression QTL): 
\begin{enumerate}[(a)] 

\item A phylogenetic tree of the 7 {\em Drosophila} species included in this study. Here we use the tree of the {\em Drosophila} 12 Genome Consortium~\cite{DrosophilaGenomesConsortium:2007gf}, which is based on genome-wide divergence at synonymous sequence sites. This tree determines six clades of phylogenetically related species (Fig.~\ref{clade_phyl}A), which are used in our analysis of time-dependent expression divergence (Figs. 1A and 2A,B). 

\item Divergence times between all pairs of species, scaled in units of the inverse neutral point mutation rate. The tree of Fig.~\ref{clade_phyl}A uses a lineage-specific mutation rate to infer the length of its 12 branches. The scaled divergence time $\tau_{ij}$ for a given species pair $(i,j)$ is the sum of the lengths of the branches connecting these species. The scaled divergence time of a clade ${\cal C}$ is defined as an average over species pairs, 
\EQ
\tau_\C = \frac{1}{|\C_1||\C \setminus \C_1|} \sum_{i \in \C_1} \sum_{j \in \C \setminus \C_1} \tau_{ij}, 
\label{tauC}
\EE
where $\C$ is the set of species in the clade and $(\C_1, \C_2)$ is the portioning of this set defined by the root node. 
These divergence times differ substantially from previous estimates based on amino acid distances~\cite{Bedford:2009fy}. 

\item The neutral nucleotide sequence diversity,
\EQ
\pi_0 =2p \mu N, 
\label{theta}
\EE
where $N$ is the effective population size, and $ p =1 ,2$ for haploid/diploid organisms, respectively. Here we use $\pi_0 = 0.0112$, as determined in ref.~\cite{Mackay:2012fd} from genome-wide polymorphism data at synonymous sequence sites. The sequence diversity enters the definition of the scaled $\Omega$ ratio in equation~\ref{Omega} and the probabilistic extension of the $\Omega$ test (section~2). 
\end{enumerate}

\paragraph{Expression data.}
We use genome-wide expression data from 7 {\em Drosophila} species obtained by Zhang et al.~\cite{Zhang:2007bf} (Gene Expression Omnibus under accession number GSE6640). These data are well suited for our analysis. They cover several clades of species that are well comparable at the organismic level and sufficiently diverged for adaptive evolution of expression to be detectable (section~2). Moreover, {\em Drosophila} has larger effective population size, higher mutation rates, and shorter generation times than typical mammalian species~\cite{Gilad:2006fv}, and adaptive evolution has been detected at the genomic level by several methods~\cite{Andolfatto:2005ir,Lassig:2007vb,Sella:2009hs}. Hence, compared to more recent data from other species~\cite{Tsankov:2010,Brawand:2011, Perry:2012}, the {\em Drosophila} expression data of Zhang et al.~\cite{Zhang:2007bf} are a suitable target for the inference of adaptive evolution. These data contain mRNA intensity measurements for a number of male and female inbred replicates in each species. Specific microarray platforms were designed for each of these species. Each platform has an array of probes mapped to assembled genome sequences and to GLEANR gene annotations by the Drosophila 12 Genomes Consortium~\cite{DrosophilaGenomesConsortium:2007gf}, which also provides sequence homology tables. We restrict the analysis to the 6332 genes that have unambiguous one-to-one orthologs across all lines and are tested by at least four probes in each microarray platform.  We obtain a set of expression levels $E^\alpha_{i, s, \kappa}$ (defined as $\log_2$ intensities) labelled by gene number $\alpha \in \{1, \dots, g \! = \! 6332\}$, species $i  \in \{ \mbox{{\em mel}, {\em sim}, {\em yak}, {\em ana}, {\em pse}, {\em vir}, {\em moj}} \}$ (Fig.~\ref{clade_phyl}), sex $s \in \{ m, f  \}$, and inbred replicates $\kappa \in\{ 1, \dots, k_{i,s} = 4-7\}$. The data contain two strains of {\em D.~simulans} (14021-0251.011  and 14021-0251.198), which are used to estimate the genetic variance of expression (see below). 
 
\paragraph{Normalization of expression levels.}
We define a linear transformation of the levels~\cite{Quackenbush:2002kl},  
\EQ 
E^\alpha_{i, s, \kappa} \to \frac{E^\alpha_{i, s, \kappa} - \langle E_{i,s,\kappa}  \rangle}{ \sqrt{V_{i,s,\kappa}}}, 
\label{norm}
\EE
where $ \langle E_{i,s,\kappa} \rangle$ and $V_{i,s,\kappa}$ denote mean and variance of the expression across all genes in a given individual $(i, s, \kappa)$. The transformed levels $E^\alpha_{i, s, \kappa}$ are shifted to mean 0 and normalized to variance 1 across all genes in each individual. The transformation (\ref{norm}) is a heuristic to reduce differences in probe sensitivity between microarrays (each individual is measured in a separate array). To test its influence on our inference of selection,  we compare the aggregate $\Omega$ ratio, which is defined in equation~(\ref{Omega}), for untransformed expression levels, expression levels with only shift ($E^\alpha_{i, s, \kappa} \to E^\alpha_{i, s, \kappa} - \langle E_{i,s,\kappa}  \rangle$), and expression levels with shift and  normalization~\cite{Quackenbush:2002kl} (Fig.~\ref{fig:rescaling}A-C). Shifting to zero mean turns out to be an essential step to remove spurious expression divergence. The subsequent normalization to variance 1 affects the $\Omega$ data and our inference of selection only weakly. However, additively and multiplicatively transformed expression levels produce less noisy $\Omega$ data than levels with only additive transformation (cf.~section~2). Hence, we use levels $E^\alpha_{i, s, \kappa}$ as given by the transformation (\ref{norm}) for our evolutionary analysis.

\paragraph{Expression statistics within and across populations.} 
Using the normalized expression levels, we can define averages and natural variation of expression at three different levels: 
\begin{enumerate}[(a)] 

\item 
The mean and variance of expression across  experimental replicates of an inbred line characterize the distribution of expression levels for a given genotype. Here we estimate these quantities from the data of each inbred line, 
\EQ
 {E}^\alpha_{i, s}  = \frac{1}{k_{i}} \sum_\kappa  {E}^\alpha_{i, s, \kappa} ,
 \qquad \qquad 
 \delta^\alpha_{i,s} = \frac{1}{k_{i} - 1}  \sum_\kappa ( {E}^\alpha_{i, s, \kappa}- {E}^\alpha_{i, s} )^2, 
\label{isogenic_statistics}
 \EE 
and we define the sample mean and variance,  
\EQ
E_i^\alpha = \frac{1}{2} ({E}^\alpha_{i, m} + {E}^\alpha_{i, f}), 
\qquad
\delta_i^\alpha = \frac{1}{2} (\delta^\alpha_{i,m} + \delta^\alpha_{i,f}). 
\label{Eia}
\EE

\item 
The genetic mean and diversity of expression characterize the distribution of heritable expression differences in a population. Heritable components of quantitative traits are often inferred from ``common garden'' breeding experiments under standardized environmental conditions. Here we estimate the genetic mean and diversity for a given gene from the data within one species, 
\EQ
\Gamma_i^\alpha = E_i^\alpha \pm  \sqrt{\frac{\Delta_i^\alpha}{2} + \frac{\delta^\alpha_{i}}{k_i}},
\qquad \qquad 
\Delta^\alpha = \frac{1}{2} (E^\alpha_{sim1} - E^\alpha_{sim2})^2 - \frac{1}{k_i} \delta^\alpha_{i},
\label{Delta}
\EE
where we have included the expected sampling error for $\Gamma_i^\alpha$. The data set of ref.~\cite{Zhang:2007bf} limits the inference of expression diversity to a broad estimate from two strains of {\em D. simulans}. This is sufficient for our analysis, because the inference of adaptation decouples from the precise value of $\langle \Delta \rangle$ (section~3). Similarly, we define the expression dimorphism between males and females in each species, 
\EQ
\Delta_{i,mf}^\alpha = \frac{1}{2} (E^\alpha_{i,m} - E^\alpha_{i,f})^2 - \frac{1}{k_i} \delta^\alpha_{i}.
\label{Delta_mf}
\EE

\item 
The expression divergence is defined as the squared difference between population means, $D^\alpha_{ij}  = (\Gamma_i^\alpha - \Gamma_j^\alpha)^2$, and characterizes evolutionary expression differences between two species. Here we estimate the divergence for a given gene from the cross-species data,  
\EQ
D^\alpha_{ij} = (E_i^\alpha - E_j^\alpha)^2 - \Delta^\alpha - \frac{1}{k_i} \delta^\alpha_{i} -\frac{1}{k_j}  \delta^\alpha_{j}. 
\label{D} 
\EE
\end{enumerate} 
Equations (\ref{Delta}) and (\ref{D}) follow Wright's decomposition of the variance of a quantitative trait into intra- and inter-species components~\cite{Wright:1950wr}, which underlies the quantitative genetics summary statistics $F_\st$ and $Q_\st$ (see section~2). For the analysis of sex-specific evolution (section~3), we use the same rationale for the sex-specificity traits $E_{i, \mf}^\alpha = E^\alpha_{i,m} - E^\alpha_{i,f}$. 

In Fig.~\ref{clade_phyl}B, we compare gene-averaged values of isogenic variance, diversity, dimorphism and divergence (these averages are denoted by angular brackets), as well as the cross-gene variance of expression. We find a clear ranking $\langle \delta_i \rangle < \langle \Delta \rangle \lesssim \langle \Delta_{i,mf} \rangle < \langle D_{ij} \rangle < V_i$ for all species $i$ and $j$, where $V_i = \langle \Gamma_i^2 \rangle \approx 1$ by our normalization. In the $\Omega$ test for selection on gene expression, we use diversity and divergence estimates given by equations (\ref{Delta}) and (\ref{D}) in aggregate measures across groups of species and classes of genes. However, our data set has a low number of individuals per species. Hence, single-gene estimates of diversity and divergence are noisy, which calls for a probabilistic inference of selection. The $\Omega$ test and its probabilistic extension for individual genes are described in section~2.

\paragraph{Divergence-diversity ratio, $\Omega$.}
The aggregate expression divergence-diversity ratio $\Omega_{ij}$ for a given species pair $(i,j)$ is defined as 
\EQA
\Omega_{ij} & =&  \pi_0\, \frac{\langle D_{ij}\rangle}{\langle \Delta\rangle}, 
\label{Omega} 
\EEA
where $\pi_0$ is the neutral sequence diversity (equation~\ref{theta}). The expression diversity and gene-specific divergence values $D^\alpha_{ij}$ and $\Delta^\alpha_{ij}$ are given by equations~(\ref{Delta}) and~(\ref{D}). Angular brackets denote averages over all genes in our dataset, $\langle D_{ij}\rangle = \frac{1}{g}\sum_{\alpha}  D^\alpha_{ij}$. The prefactor in equation (\ref{Omega}) is chosen such that $\Omega \simeq 1$ for neutral evolution in the limit of long divergence times (section~2). The divergence-diversity ratio $\Omega_\C$ for a species clade $\C$ is defined as an average over species pairs,  
\EQ
 \Omega_\C  = \frac{1}{|\C_1||\C \setminus \C_1|} \sum_{i \in \C_1} \sum_{j \in \C \setminus \C_1} \Omega_{ij},  
\label{OmegaC}
\EE
in analogy with the definition (\ref{tauC}) of clade divergence times. We also define divergence-diversity ratios $\Omega_{ij}^{\cal G}$ and $\Omega^{\cal G}_\C$ for specific gene classes ${\cal G}$, using restricted averages $\langle \dots \rangle_{\cal G}$.

\subsubsection*{2. Inference of selection on gene expression}

\paragraph{Evolutionary model.}
We consider the evolution of gene expression levels under genetic drift, mutation, and selection given by a fitness model with peak displacements on macro-evolutionary time scales. In the minimal seascape model~\cite{Held:2014ve, NourMohammad:2013in}, the fitness of a given gene depends on its expression level $E$ and on evolutionary time $t$,
\EQ
f(E,t) = f^* - c_0 \big (E - E^*(t) \big )^2. 
\label{fitness}
\EE
The expression value of maximum fitness, $E^*(t)$, performs an Ornstein-Uhlenbeck random walk with diffusion constant $\vv_0$, average value $\cE$ and stationary mean square deviation $r^2E_0^2$, where $r^2$ is a constant of order 1. This process is defined by the Langevin equation 
\EQ
\frac{d}{dt} E^*(t) =   -\frac{\vv_0}{2r^2 E_0^2} (E^*(t) - \cE) + \eta(t) , 
\label{eq:langEstar}
\EE 
where $\eta(t)$ is the random variable of a delta-correlated Gaussian process with average 0 and variance $\vv_0$. These random variables are assumed to be independent for each gene and on each lineage. The Ornstein-Uhlenbeck fitness seascape should not be confused with a previous Ornstein-Uhlenbeck model for the evolution of quantitative traits  under stabilizing selection~\cite{Hansen:1997ws,Butler:2004th,Hansen:2008gt,Bedford:2009fy,Kalinka:2010dw,Beaulieu:2012ex,Rohlfs:2014bl} (a detailed comparison is given below).  

The minimal seascape  model captures two kinds of selection on gene expression in a unified way:
\begin{enumerate}[(a)] 

\item Stabilizing selection. This type of selection constrains the intra- and inter-population variation of expression levels to values around $E^*(t)$. We define the dimensionless {\em stabilizing strength}
\EQ
c = 2 N\, E_0^2 \,c_0, 
\label{c}
\EE
where $N$ is the effective population size and the trait scale $E_0^2$ is given by the average genetic variation of expression in the long-term limit of neutral evolution~\cite{Nourmohammad:2013ty}, $E_0^2 = \lim_{\tau \to \infty} \langle D (\tau) \rangle /2$. In the limit case $\vv_0 = 0$, the fitness seascape reduces to a static fitness landscape, $f(E) = f^* - c_0 (E - E^*)^2$, and stabilizing selection is the only selective force. This provides a simple interpretation of the selection parameter $c$: it compares the (hypothetical) genetic load $c_0 \,E_0^2$ of a neutrally evolving trait evaluated in the landscape $f(E)$ and the actual genetic load $1/2N$ in the same landscape, assuming a mutation-selection-drift equilibrium at low mutation rates~\cite{NourMohammad:2013in}. This parameter signals the regimes of weak ($c \lesssim 1$) and strong ($c \gtrsim 1$) stabilizing selection~\cite{Nourmohammad:2013ty}. 

\item Directional selection. In a fitness seascape,  this type of selection triggers adaptive response of the population mean trait in the direction of fitness peak displacements. We define the scaled {\em driving rate} 
\EQ
\vv = \frac{\vv_0}{\mu \, E_0^2}. 
\label{v}
\EE
This parameter measures mean square displacement of the fitness peak, in units of $E_0^2$ and per unit $1/\mu$ of evolutionary time. In {\em macro-evolutionary} seascapes, $\vv$ is sufficiently low for population to follow fitness peak displacements; such seascapes are a joint model of stabilizing and directional selection~\cite{Held:2014ve}. The values of $\vv$ inferred from our data fall in this regime (see section~2).  Because the seascape dynamics is a short-range Markov process, the mean square peak displacement over a scaled evolutionary time $\tau$ is then simply $E_0^2\, \vv \tau$. (Here we express $\vv$ in units of $\mu$ and $\tau$ in units of $1/\mu$, which differs slightly from the notation in refs.~\cite{Held:2014ve, NourMohammad:2013in}.) In the long-term regime $ \vv \tau \gg r^2$, the fitness peak dynamics becomes stationary with mean $\cE$ and variance $r^2 E_0^2$. This regime turns out to be well beyond the divergence times in our species sample. Hence, the statistics of {\em Drosophila} gene expression levels and our inference of selection are independent of $r^2$. 

\end{enumerate}

\paragraph{Fitness flux.} 
This measure of adaptation is defined as the speed of movement on a fitness land- or seascape by genotype or heritable phenotype changes in a population~\cite{Mustonen:2010ig, Held:2014ve}. The cumulative fitness flux associated with the population mean expression level $\Gamma (t)$ of a gene in a fitness seascape $f(E,t)$ is given by 
\EQ
\Phi (\tau) = \int_{t=0}^{\tau} \frac{\partial f(\Gamma, t)}{\partial \Gamma} \, \frac{\d \Gamma (t)}{ \d t} \, \d t . 
\label{Phidef}
\EE
This quantity measures the total amount of adaptation over a macro-evolutionary period $\tau$ in a population history. This quantity satisfies the fitness flux theorem~\cite{Mustonen:2010ig}, which generalizes the Fisher's fundamental theorem of natural selection to mutation-selection-drift processes. As shown by the fitness flux theorem, the average cumulative fitness flux over parallel evolutionary histories, in units of $1/2N$, measures the importance of adaptation compared to genetic drift: adaptation is substantial if $\langle 2 N \Phi (\tau) \rangle \gtrsim 1$. For a stationary adaptive process in the minimal seascape (\ref{fitness}), the average scaled cumulative fitness flux takes the simple form~\cite{NourMohammad:2013in,Held:2014ve}
\EQ
\langle 2 N\Phi (\tau) \rangle \simeq  2c \vv \, \tau, 
\label{Phicv}
\EE
up to factors of order $\pi_0$, as quoted in Box~1.  The exact functional form  of the fitness flux is given in reference~\cite{Held:2014ve}. This quantity is closely related to the time-dependent fraction of expression divergence that is adaptive, $\omega_\ad (\tau)$ (equation~\ref{omegaPhi}). We introduce the shorthand $\Phi =\Phi (\tau_{\rm Dros.})$ with $\tau_{\rm Dros.} = 1.4$ (Fig.~\ref{clade_phyl}A); this quantity measures the amount of adaptation across the {\em Drosophila} genus. By the probabilistic inference method discussed below, we obtain expectation values $2N\Phi^\alpha$ of the rescaled fitness flux for individual genes over the divergence time of the {\em Drosophila} genus (equation~\ref{Phi_alpha}). We use these values to describe the overall statistics of expression adaptation (Fig.~1B), to infer differences in adaptation between gene classes (Fig.~2B,D; Table~\ref{table:fraction}), and to define significantly adaptive genes (using a threshold $2N \Phi^\alpha > 4$; Table~S1). For the analysis of sex-specific adaptation (Fig.~2C,D), we define an analogous fitness flux $2N\Phi_\mf$ for sex-specificity traits (section~4).

\paragraph{Evolutionary modes of quantitative traits.} 
In the minimal seascape model, the aggregate $\Omega$ ratio defined by equation (\ref{Omega}) depends on the divergence time $\tau$ and on the selection parameters $c$ and $\vv$. We can use this dependence to distinguish three modes of evolution~\cite{Held:2014ve, NourMohammad:2013in}: 
\begin{enumerate}[(a)] 

\item {Neutral evolution} ($c = 0$). The divergence-diversity ratio has an initially linear increase due to mutations and genetic drift, and it approaches a maximum value~1 with a scaled relaxation time of~1,  
\EQ
\Omega_0 (\tau)  \simeq 
\left \{
\begin{array}{ll}
\tau   & \mbox{ for $\tau \ll 1,$} 
\\
1 & \mbox{ for $\tau \gg 1.$}
\end{array} \right. 
\label{Omega_neutral}
\EE
The short-term behavior reflects the linear growth of the average divergence~\cite{Chakraborty:1982wp,Lynch:1986tu,Lynch:1998vx},  
\EQ
\langle D(\tau)\rangle_0 \simeq  \frac{\langle \Delta \rangle_0}{\pi_0} \, \tau
\qquad \mbox{for $\tau \ll 1.$}
\label{D0Delta0}
\EE
The growth rate is the average diversity at neutrality divided by the sequence diversity (equation~\ref{theta}). This ratio equals the mutational variance of a quantitative trait as defined in refs.~\cite{Chakraborty:1982wp,Lynch:1986tu,Lynch:1998vx}, up to a rescaling of evolutionary time to units of $1/\mu$. 

\item{Evolution under stabilizing selection} ($c \gtrsim 1, \vv = 0$).
In a static fitness landscape, the divergence-diversity ratio approaches a smaller maximum value, $\Omega_{\rm stab} (c) < 1$, with a proportionally shorter relaxation time~\cite{Held:2014ve}, 
\EQ
\Omega_\eq (\tau) \simeq
\left \{
\begin{array}{ll}
\tau  & \mbox{ for $\tau \ll \Omega_{\rm stab} (c)$} 
\\
\Omega_{\rm stab} (c) & \mbox{ for $\tau \gg \Omega_{\rm stab} (c).$}
\end{array} \right.   
\label{Omega_cons}
\EE
Over a wide range of evolutionary parameters, which includes the inferred values for the data set of this study, the maximum value depends on the stabilizing strength in a simple way, $\Omega_{\rm stab} (c) \sim 1/(2c)$, with corrections for weaker
selection and for larger nucleotide sequence diversity~\cite{Nourmohammad:2013ty}.

\item{Adaptive evolution under stabilizing and directional selection} ($c \gtrsim 1, \vv > 0$).
In a genuine fitness seascape, the divergence-diversity ratio acquires an adaptive component, 
\EQ
\Omega (\tau)  =  \Omega_\eq (\tau) + \Omega_\ad (\tau)  
=  \left \{ \begin{array}{ll}
\tau  & \mbox{ for $\tau \ll \Omega_{\rm stab} (c)$} 
\\
\Omega_{\rm stab} (c)  + \frac{1}{2} \vv \,[\tau- 2 \Omega_{\rm stab} (c)], & \mbox{ for $\tau \gg \Omega_{\rm stab} (c),$}
\end{array} \right.     
\label{Omega_ad}
\EE
with corrections for $\tau$ approaching the saturation time of fitness peak displacements, $r^2/ \vv$. The universal short-term behavior $\Omega (\tau) \simeq \tau$ (equations~\ref{Omega_cons} and~\ref{Omega_ad}) reflects the quasi-neutral growth of the divergence~\cite{NourMohammad:2013in,Held:2014ve}, 
\EQ
\langle D(\tau)\rangle \simeq  \frac{\langle \Delta \rangle (c)}{\pi_0} \, \tau
\qquad \mbox{ for $\tau \ll \Omega_{\rm stab} (c)$,}
\label{qn}
\EE
where $\langle \Delta \rangle (c)$ is the average diversity under selection. As shown by comparison with the neutral behavior (equation~\ref{D0Delta0}), selection enters only via the constraint on $\langle \Delta \rangle$. Over a wide range of the stabilizing strength $c$, this constraint remains weak and $D(\tau)$ evolves near neutrality~\cite{Nourmohammad:2013ty}, as long as $\tau \ll \Omega_{\rm stab} (c)$.
\end{enumerate}
The full analytical form of the functions $\Omega_0 (\tau)$ (equation~\ref{Omega_neutral}), $\Omega_\eq (\tau)$ (equation~\ref{Omega_cons}), and $\Omega (\tau)$ (equation~\ref{Omega_ad}) is given in refs.~\cite{Held:2014ve, NourMohammad:2013in}.

\paragraph{The $\Omega$ test for selection.} 
The evolutionary statistics of the $\Omega$ ratio provides a joint test for stabilizing and directional selection on quantitative traits. 
We can infer the selection parameters of a seascape model by fitting the function $\Omega (\tau)$ (equation~\ref{Omega_ad}) to data $(\tau, \Omega)$. This method has the following properties:
\begin{enumerate}[(a)]

\item The $\Omega$ test requires data $(\tau, \Omega)$ from species with different divergence times in the regime $\tau \gtrsim \Omega_\stab$. In the quasi-neutral regime $\tau \lesssim \Omega_\stab$, the $\Omega$ ratio is insensitive to selection (equations~\ref{Omega_neutral}--\ref{Omega_ad}). 

\item By the decomposition (equation~\ref{Omega_ad}), the $\Omega$ test infers a time-dependent fraction 
\EQ
\omega_\ad (\tau) = \frac{\Omega_\ad (\tau)}{\Omega (\tau)} 
\label{adaptive_divergence_aggregate}
\EE
of the aggregate trait divergence to be adaptive. The complementary fraction, $1 - \omega_\ad (\tau)$, is attributed to genetic drift under stabilizing selection. 

\item We can approximate the divergence-diversity ratio (equation~\ref{Omega_ad}) by the linear form $\Omega(\tau) \approx \Omega_{\rm stab} + \Omega_\ad (\tau) =  \Omega_{\rm stab} + \vv \tau /2$. Therefore, already a linear fit to data produces simple estimates of stabilizing strength and driving rate, 
\EQ
c \approx \frac{1}{2 \Omega_\stab}, 
\qquad 
\vv \approx \frac{2 \Omega_\ad (\tau)}{\tau} ,
\label{cv_inferred}
\EE
and infers the adaptive fraction of expression divergence, which is related to the average scaled fitness flux~\cite{Held:2014ve} (equation~\ref{Phicv}), 
\EQ
\omega_\ad (\tau) \approx \frac{\Omega (\tau) - \Omega_\stab}{\Omega (\tau)},
\qquad
\langle 2N\Phi (\tau) \rangle \approx  \frac{2\omega_\ad (\tau)}{1 - \omega_\ad (\tau)} \approx \frac{2\big(\Omega (\tau) - \Omega_\stab\big)}{\Omega_\stab}. 
\label{omegaPhi}
\EE

\item The $\Omega$ ratio of a quantitative trait depends on the selection parameters $c$ and $\vv$, but it decouples from the genetic basis of the trait. Specifically, it depends only weakly on the number and effect size of the underlying QTL~\cite{Nourmohammad:2013ty,Held:2014ve}, on the amount of recombination between these sites~\cite{Nourmohammad:2013ty,Held:2014ve}, and on the nonlinearities in the genotype-phenotype map (trait epistasis; see section 5 and Fig.~\ref{fig:sim}B). The  $\Omega$ statistics also decouples from details of the selection dynamics; it can be applied to {\em punctuated} adaptive processes, which have fewer and larger peak displacements~\cite{Held:2014ve} (section~3).  
\end{enumerate}

\paragraph{Application of the $\Omega$ test to gene expression data.} 
In Fig.~1, we compare the model function $\Omega (\tau)$ (equation~\ref{Omega_ad}) to aggregate expression data $(\tau_\C, \Omega_\C)$ for six {\em Drosophila} species clades (equations~\ref{tauC} and~\ref{OmegaC}). The best-fit seascape model ($c^* = 18.6$, $\vv^* = 0.07$; green line) explains these data, which produces evidence for adaptive evolution of gene expression. Using the decomposition into adaptive and drift components (green and blue shaded areas), we obtain a cumulative fitness flux $\langle 2N\Phi \rangle = 3.6$ across the entire {\em Drosophila} genus (equations~\ref{cv_inferred} and~\ref{omegaPhi}). Importantly, the inference of adaptive evolution decouples from the precise overall scale of $\Omega$, which is influenced by our limited information on expression diversity (section~2).

\paragraph{Control analysis of equilibrium models.} 
We can also compare the aggregate expression data $(\tau_\C, \Omega_\C)$ to models of time-independent stabilizing selection: 
\begin{enumerate}[(a)]
\item Fitness landscape model. In contrast to the seascape model, the best-fit landscape model ($c_\eq =13$, $\vv = 0$) provides a poor fit to the data (Fig.~\ref{fig:Hartl}A). It captures the average $\Omega$ ratio across the {\em Drosophila} clades, but fails to describe the systematic amplitude differences between these clades. In particular, the landscape model drastically overestimates the divergence of close species, $D_{ {mel}-{yak}}$ and $D_{ {mel}-{sim}}$. 

\item Ornstein-Uhlenbeck model. In a previous study, Bedford and Hartl~\cite{Bedford:2009fy} analyze  the same data set and infer broad stabilizing selection on expression levels, which is consistent with our results. However, they observe a saturation of gene expression divergence that is at variance with the inference of a linear growth on time scales beyond the divergence time of {\em D. melanogaster} and {\em D. simulans} (Fig.~1A and ref.~\cite{Zhang:2007bf}). This can be traced to differences in data analysis. First, Bedford and Hartl~\cite{Bedford:2009fy} use amino acid distances in their phylogeny. These distances are affected by selection~\cite{Smith:2002cm} and produce relative branch lengths that differ substantially from the phylogeny based on fourfold synonymous sites~\cite{DrosophilaGenomesConsortium:2007gf} used in this study (Fig.~\ref{clade_phyl}A and Fig.~\ref{fig:Hartl}). Second, Bedford and Hartl~\cite{Bedford:2009fy} analyze expression divergence for pairs of species, while we group the species into clades (Fig.~\ref{clade_phyl}A). These differences lead to a more noisy dependence of the expression divergence data on evolutionary time and make the distinction of conservation and adaptation more difficult (Fig.~\ref{fig:Hartl}B). Bedford and Hartl~\cite{Bedford:2009fy} fit these data to an Ornstein-Uhlenbeck model of evolution under stabilizing selection~\cite{Hansen:1997ws} (equation~\ref{ou}), which is described below. This model has two independent  parameters, which equals the number of fit parameters for our minimal seascape model. Similarly to our landscape model, the best-fit Ornstein-Uhlenbeck model explains the average expression divergence across the {\em Drosophila} genus (Fig.~\ref{fig:Hartl}B), but it cannot explain the pattern of expression divergence between close species. The model predicts a quasi-neutral linear growth of the divergence with $D_{ {mel}-{yak}} / D_{ {mel}-{sim}} \approx \tau_{ {mel}-{yak}} / \tau_{ {mel}-{sim}} \approx 2$ (equation~\ref{qn}), which drastically overestimates the observed ratio  $D_{ {mel}-{yak}} / D_{ {mel}-{sim}} \approx 1.2$. 
\end{enumerate}

\paragraph{Comparison of the $\Omega$ test with related methods.}
Our inference method for selection on quantitative traits can be compared with three well-known selection tests for phenotypic and genomic data: 
\begin{enumerate}[(a)]

\item $Q_\st / F_\st$ ratio test for selection on quantitative traits. The summary statistics $F_\st$ and $Q_\st$ measure the expected fraction of the total genetic variation harboured in a pair of populations that can be attributed to the divergence between these populations; the complementary fraction is attributed to the diversity within populations. $F_\st$ refers to neutrally evolving sequence loci~\cite{Wright:1943wy,Wright:1950wr,Lande:1992vo}, which can be regarded as a ``pseudotrait'' with aggregate divergence and diversity. $Q_\st$ is the analogous measure for quantitative traits under selection~\cite{Spitze:1993vo}. The expected dependence of these measures on divergence time can be expressed in terms of the $\Omega$ ratio (equation~\ref{Omega}), 
\EQA
F_\st (\tau) & = & \frac{\langle D (\tau) \rangle_0}{\langle D (\tau) \rangle_0 + 2 \langle \Delta \rangle_0} = \frac{\Omega_0 (\tau)}{\Omega_0 (\tau) + 2\pi_0}, 
\label{Fst}
\\
Q_\st (\tau) & = & \frac{\langle D (\tau) \rangle}{\langle D (\tau) \rangle + 2 \langle \Delta \rangle}  = \frac{\Omega (\tau)}{\Omega (\tau) + 2\pi_0}, 
\label{Qst}
\EEA
where we use expectation values $\langle \dots \rangle$ in an ensemble of parallel-evolving populations and the subscript 0 refers to neutral evolution. The $Q_\st/F_\st$ test~\cite{Leinonen:2013ht} stipulates that a quantitative trait is evolving at neutrality if $Q_\st/F_\st = 1$, under stabilizing selection if $Q_\st/F_\st < 1$, and under directional selection if $Q_\st/F_\st > 1$. Comparison with the theory of the $\Omega$ ratio (equations~\ref{Omega_neutral}--\ref{qn}) shows that the $Q_\st/F_\st$ test is insensitive to selection for divergence times in the quasineutral regime,
\EQ
\frac{Q_\st (\tau)}{F_\st (\tau)}\simeq 1
\qquad \mbox{for $\tau \ll \Omega_\stab (c)$.} 
\label{qstfst}
\EE
The data set of this study, which has divergence times $\tau \geq \Omega_\stab$ and aggregate values $Q_\st/F_\st$ between $0.6$ for the {\em mel}-{\em sim} clade and $0.8$ across the entire {\em Drosophila} genus; these values are obtained using equations~(\ref{OmegaC}), (\ref{Fst}), and (\ref{Qst}). Hence, this test signals broad stabilizing selection but no directional selection. In contrast, the $\Omega$ test infers both stabilizing and directional selection from the linear dependence $\Omega (\tau)$ (Fig.~1A and equation~\ref{Omega_ad}). This inference shows a conceptually important point: stabilizing and directional selection are not mutually exclusive, but joint features of selection on macro-evolutionary time scales.

\item Ornstein-Uhlenbeck model for quantitative trait evolution. This phenomenological model describes a quantitative trait evolving under genetic drift and stabilizing selection~\cite{Hansen:1997ws,Butler:2004th,Hansen:2008gt,Beaulieu:2012ex} and has been applied to the evolution of gene expression~\cite{Bedford:2009fy, Kalinka:2010dw,Rohlfs:2014bl} (a detailed comparison with the results of ref.~\cite{Bedford:2009fy} is given above). The model is defined by a Langevin equation for the population mean trait, 
\EQ
\frac{d}{dt} \Gamma (t) = - \lambda \, (\Gamma - E^*) + \eta_\Gamma (t),
\label{ou}
\EE
where $\eta_\Gamma (t)$ is the random variable of a delta-correlated Gaussian process with average 0 and variance $\sigma^2$. 
The model constants $\lambda$ and $\sigma^2$ are usually regarded as independent fit parameters. 
The Ornstein-Uhlenbeck dynamics of the population mean trait $\Gamma (t)$ around a fixed optimal trait value $E^*$ (equation \ref{ou}) should not be confused with the Ornstein-Uhlenbeck dynamics of the time-dependent optimum $E^*(t)$ in our seascape model (equation~\ref{fitness}). A Langevin equation similar to (\ref{ou}) can be derived from a more general population-genetic model for the evolution of a quantitative trait $E$ a static fitness landscape $f(E) = -c_0 \, (E - E^*)^2$, which has been introduced in ref.~\cite{deVladar:2011bs,Nourmohammad:2013ty}. In this model, the population mean trait follows the Ornstein-Uhlenbeck process 
\EQ
\frac{d}{dt} \Gamma (t) =  - 2 \langle \Delta \rangle \, c_0 \, (\Gamma - E^*) - 2 \mu \, (\Gamma - \Gamma_0) + \eta_\Gamma (t),
\label{ou2}
\EE
where $\Gamma_0$ is the genetic mean trait in the long-term limit of neutral evolution and $\eta_\Gamma (t)$ is the random variable of a delta-correlated Gaussian process with average 0 and variance $\langle \Delta \rangle / N$. Comparison with equation (\ref{ou}) determines the Ornstein-Uhlenbeck coefficients in terms of the stabilizing strength and the average trait diversity ($\lambda = 2 \langle \Delta \rangle \, c_0$, $\sigma^2 = \langle \Delta \rangle / N$). Equation (\ref{ou2}) contains an additional mutational  term $(-2 \mu) (\Gamma - \Gamma_0)$, which implies that the expectation value $\langle \Gamma \rangle$ differs from the optimum trait value $E^*$. We note that the diffusion constant $\langle \Delta \rangle / N$ determines the behavior of the $\Omega $ ratio (equation~\ref{Omega_cons}), of the trait divergence (equation~\ref{qn}), and of the $Q_\st / F_\st$ ratio (equation~\ref{qstfst}) in the quasineutral regime ($\tau \ll \Omega_\stab$). The Ornstein-Uhlenbeck model has been generalized to account for lineage-specific stabilizing selection in a phylogeny~\cite{Hansen:1997ws,Butler:2004th,Hansen:2008gt,Kalinka:2010dw,Beaulieu:2012ex,Rohlfs:2014bl}; however, inferring independent selection parameters for each lineage may lead to overfitting of our data set. 
Instead, we use the seascape model (\ref{fitness}) to infer lineage- and gene-specific changes of the trait optimum $E^*(t)$ using a single additional selection parameter $\vv$.

\item McDonald-Kreitman test for adaptive sequence evolution~\cite{Kreitman:1991vh}. The conceptually closest sequence-based test evaluates the sequence divergence-diversity ratio $\Omega$ for a sequence class under putative selection (e.g., non-synonymous mutations in protein-coding sequence)  and compares it to the analogous ratio $\Omega_0$ for {\em bona fide} neutral changes (e.g., synonymous mutations). Positive selection in the query sequence is inferred if $\Omega > \Omega_0$. In this case, the amplitude ratio $\alpha_{\rm seq} = (\Omega - \Omega_0)/\Omega$ estimates the fraction of non-synonymous substitutions that are adaptive, i.e., driven by positive selection~\cite{Smith:2002cm}. This is a variant of the McDonald-Kreitman test~\cite{Kreitman:1991vh}. It requires data from query sequence and from neutral sequence, but only from a single pair of species with a divergence time beyond the coalescence time. In contrast, the $\Omega$ test requires only data from traits under selection, but from three or more species with divergence times beyond the equilibrium relaxation time $\Omega_\stab$. These differences highlight distinct evolutionary characteristics of quantitative traits. First, such traits have a quasi-neutral regime of  macro-evolutionary divergence times (equation~\ref{qn}) that has no direct analogue in sequence evolution~\cite{Nourmohammad:2013ty}. Second, in most cases we do not have a gauge of neutrally evolving traits analogous to synonymous sequence. 
\end{enumerate}

\paragraph{Probabilistic inference of selection.}
Here we describe the extension of our selection inference method to expression data of individual genes. A minimal seascape model is determined by the parameters $(c, \vv)$ or equivalently by $(c, \Phi)$, where $\Phi =2 c \vv \tau_{\rm Dros.} / 2N$ denotes the expected cumulative fitness flux over the genus divergence time (equation~\ref{Phicv}). We derive a posterior probability distribution $Q(c,\Phi \, | \, \E^\alpha)$, where $\E = (E^\alpha_1, \dots, E^\alpha_7)$ denotes the expression levels of gene $\alpha$ in the 7 species of our data set. This derivation consists of three steps: we obtain the probability distribution $Q(\GGamma \, | \, c, \Phi)$ of population mean traits $\GGamma^\alpha = (\Gamma^\alpha_1, \dots, \Gamma^\alpha_7$) in a given seascape model, we include sampling effects to determine the distribution $Q(\E  \, | \, \Phi)$, and we use Bayes' theorem to infer the posterior distribution $Q(c, \Phi \, | \, \E^\alpha)$. 

The basic building block of evolutionary statistics in the minimal seascape model has been derived previously~\cite{Held:2014ve}: the {\em lineage propagator} $G_\tau (\Gamma, E^* | \Gamma_a, E^*_a)$ is the probability density of mean and optimal trait values $(\Gamma, E^*)$, given the values $(\Gamma_a, E^*_a)$ in an ancestral population at scaled evolutionary distance $\tau$. The lineage propagator is related to the stationary distribution of the seascape dynamics, $Q_{\rm stat} (\Gamma, E^*) = \lim_{\tau \to \infty} G_\tau (\Gamma, E^* | \Gamma_a, E^*_a)$. Both distributions are Gaussian functions that depend on the seascape model parameters and on the neutral variance (trait scale) $E_0^2$; their detailed analytical form is given in equations (30)--(33) and (A.15)--(A.20) of ref.~\cite{Held:2014ve}. The probability distribution of population mean traits across the {\em Drosophila} genus is the stationary distribution for its last common ancestor multiplied by the lineage propagators for all branches of the phylogeny; this expression is to be integrated over all unknown expression levels. Specifically, we obtain 
\EQ
Q(\GGamma^\alpha \, | \, c, \Phi, {E_0^2}) = 
\int \, Q_{\rm stat} (\Gamma^\alpha_{l}, E_{l}^*) \, \prod_{i=1}^{l-1} G_{\tau (i)} (\Gamma^\alpha_i, E_i^* \, | \,  \Gamma^\alpha_{a(i)}, E_{a(i)}^*) 
\, \d \Gamma^\alpha_{k+1} \dots \d \Gamma^\alpha_{l} \, 
\d E^*_1 \dots \d E_{l} ,  
\label{QGG}
\EE
where $i = 1, \dots, k$ labels the extant species and $i = k+1, \dots, l$ the clade ancestor species (with $l= 2k-1 = 13$ and the index $l$ referring to the last common ancestor of all species), $a(i)$ denotes the closest ancestor of species $i$, and $\tau (i)$ is the scaled length of the branch between $i$ and $a(i)$. The deviations of the expression measurements $E^\alpha_i$ from the population mean trait $\Gamma^\alpha_i$ can be described by a Gaussian sampling error model with variance $(\Delta^\alpha_i/2) + (\delta^\alpha_i /k_i)$, as given by equation (\ref{Delta}). We obtain  
\EQ
Q(\E^\alpha \, | \, c, \Phi, E_0^2) = 
\frac{1}{Z} \int Q(\GGamma^\alpha \, | \, c, \Phi) \, \exp \left [ - \frac{1}{2} \frac{(E^\alpha_i - \Gamma^\alpha_i)^2}{(\Delta^\alpha_i/2) + (\delta^\alpha_i /k_i)} \right ] \, \d \Gamma^\alpha_{1} \dots \d \Gamma^\alpha_{k} ,
\label{QEE}
\EE
where $Z$ is a normalization factor. This multi-variate Gaussian integral can be evaluated in a straightforward way by the saddle point method. Here we approximate the noisy diversity values of individual genes by the species averages $\Delta_i$ and $\delta_i$.   Finally, Bayes' theorem gives the posterior distribution
\EQ
Q(c, \Phi \, | \, \E^\alpha) = \frac{ \int Q(\E^\alpha \, | \, c, \Phi) \, P_0 (c, \Phi) \, {\d E_0^2}}{ \int Q(\E^\alpha \, | \, c, \Phi) \, P_0 (c, \Phi) \, {\d E_0^2} \, \d c \, \d \Phi } ,
\label{Qpost}
\EE
where $P_0 (c, \Phi)$ denotes the prior distribution of seascape parameters. This distribution determines the maximum likelihood posterior values of stabilizing strength, fitness flux, and adaptive fraction of expression divergence, 
\EQ
(c^\alpha ,\Phi^\alpha ) =  \arg \max_{c,\Phi}   Q(c, \Phi \, | \, \E^\alpha) , 
\qquad
\omega^\alpha_\ad (\tau) = \frac{(\tau / \tau_{\rm Dros.}) \, \Phi^\alpha }{ (\tau / \tau_{\rm Dros.}) \, \Phi^\alpha + 1/N} \; ; 
\label{Phi_alpha}
\EE
see equation (\ref{omegaPhi}). In equation (\ref{Qpost}), we use a prior distribution $P_0 (c, \Phi) \sim \exp (- a c - b \Phi )$ with Lagrange multipliers $a,b$ that calibrate the average posterior values $\langle c \rangle$ and $\langle  \Phi \rangle$ over all genes to our inference from aggregate data (see above). This choice generates a conservative inference of gene-specific seascape parameters that reflects two statistical features of our data. First, gene data $\E$ explained by a seascape model with parameters $(c, \Phi)$ and a neutral trait variance $E_0^2$ (see text above and ref.~\cite{Nourmohammad:2013ty}) have a similar likelihood in a family of models with parameters $(\lambda c, \Phi)$ and neutral trait variance $\lambda E_0^2$, where $\lambda > 0$ is a rescaling factor, as long as the stabilizing strength is above some minimum value. {In other words, there is a residual freedom in model parameters that leaves the fitness flux $\Phi$ invariant.} This freedom exists because the gene-specific diversity values $\Delta^\alpha_i$ are too noisy to be included in the inference. Our prior distribution favors posterior values $c$ close to the minimum stabilizing strength, which are consistent with the inference from aggregate data. Second, the distribution (\ref{QGG}) has an algebraic tail, $Q(\E \, | \, c, \Phi) \sim \Phi^{-1}$ for $2N\Phi \gg 1$, which is caused by the diffusive dynamics of the fitness peak. Our prior distribution suppresses this tail and favors posterior values $\Phi$ close to the maximum-likelihood value $\Phi^*$. The validation of this inference scheme by simulation tests is described in section 5.

\paragraph{Statistical significance of the inference.}
The probabilistic extension of the $\Omega$ plays an important role in our global inference: to quantify the statistical significance of our evidence for adaptive evolution under directional selection. Specifically, we evaluate the cumulative log-likelihood score for all genes of our data set as a function of the evolutionary variables $c$ and $\Phi$,
\EQ
S (c, \Phi) = \sum_{\alpha = 1}^g \log Q( c, \Phi \, | \, \E^\alpha),
\EE
where $Q( c, \Phi \, | \, \E^\alpha)$ is given by equation (\ref{Qpost}). This function is shown in Fig.~1C with its maximum shifted to 0. The global maximum-likelihood seascape model has parameters 
\EQ
(c^*, \Phi^*) = \arg \max_{c, \Phi} S(c, \Phi)  = \left (18.6, \frac{3.6}{2N} \right), 
\label{ML_c_phi}
\EE
compared to $(c_\eq^*, 0) = \arg \max_{c} S(c, \Phi \! = \! 0) = (16, 0)$ for the best landscape model and $(c, \Phi) = (0,0)$ for neutral evolution.  We obtain log-likelihood differences $S(c^*, \Phi^*) - S(c_\eq^*, 0) = 8396$ and $S(c^*, \Phi^*) - S(0, 0) =12464$. By a log-likelihood test, these differences translate into the $P$ values quoted in the main text. Maximum-likelihood values analogous to equation (\ref{ML_c_phi}) can also be defined for classes of genes (Table~\ref{table:fraction}).

\subsubsection*{3. Analysis of alternative evolutionary scenarios} 

\paragraph{Lineage-specific demography.} Demographic effects, such as population bottlenecks, affect the patterns of sequence variation in Drosophila~\cite{Lachaise:1988tg,Glinka:2003tr,Stephan:2007gc,Thornton:2007bl,Aquadro:2001wb}. Here we examine the effects of strong, long-term demographic heterogeneities on the divergence and diversity of expression levels. Specifically, we consider changes in effective population size to a value $N_i = \lambda N$ in a given {\em Drosophila} lineage $i$, which is defined by the terminal branch of species $i$ in the phylogeny and extends over a scaled evolutionary time $\tau_i$ (Fig.~\ref{clade_phyl}A). A depletion of effective population size leads to a global relaxation of stabilizing selection on gene expression, given by a reduced stabilizing strength $\lambda c$ in the fitness seascape (equation \ref{fitness}). For each clade $\C$ with $i \in \C$, we define the polarized divergence-diversity ratio, 

\EQ
\Omega_{\C, i} =  \frac{1}{|\C \setminus \C_1|} \sum_{j \in \C \setminus \C_1} \Omega_{ij},  
\label{OmegaCi}
\EE
where $(\C_1, \C \setminus \C_1)$ is the partioning of clade $\C$ defined by its root node and we assume $i \in \C_1$.  The pairwise ratios $\Omega_{ij}$ are given by equation (\ref{Omega}). Similarly, we define the polarized divergence time, 
\EQ
\tau_{\C, i} =  \frac{1}{|\C \setminus \C_1|} \sum_{j \in \C \setminus \C_1} \tau_{ij}.
\label{TauCi}
\EE
In Fig.~\ref{fig:demographic}A,B, we plot polarized data $(\tau_{\C,i}, \Omega_{\C,i})$ together with background data $(\tau_\C, \Omega_\C)$ from partial clades excluding species $i$. Under a change of population size in lineage $i$ with $\tau_i \gtrsim \Omega_\stab (\lambda c)$, the polarized data should follow a pattern with reduced ($\lambda < 1$) or increased ($\lambda > 1$) long-term constraint,
\EQA
\Omega (\tau, \tau_i)  & =  & \Omega_\eq (\tau, \tau_i) + \Omega_\ad (\tau, \tau_i)  
\label{Omega_dem} \\
& = & \left \{ \begin{array}{ll}
\tau  & \mbox{ for $\tau \ll \Omega_{\rm stab} (\lambda c)$} 
\\
\\
\frac{1}{2} (\Omega_{\rm stab} (\lambda c) + \Omega_\stab (c) )+ \frac{1}{2} \vv \tau+ \mathcal{F}(\lambda, c)
& \mbox{ for $\tau \gg \tau_i + \Omega_{\rm stab} (c),$}
\end{array} \right.     
\nonumber
\EEA
where the shift $\mathcal{F}(\lambda, c)$ is generated by the demographic inhomegeneity on intermediate time scales; this pattern is shown in Fig.~\ref{fig:demographic}A,B for $\lambda = 1/2$ and $\lambda=3$. A similar calculation shows that short-term population bottlenecks have a negligible effect on the $\Omega$ statistics. We observe no deviation between polarized and background $\Omega$ data, indicating the absence of strong demographic effects shaping the evolution of expression levels. Equation (\ref{Omega_dem}) also shows that demographic effects do not confound the $\Omega$ test for  adaptive evolution under directional selection. For time-independent optimal trait value ($\vv = 0$), global relaxation of stabilizing selection increases the divergence-diversity ratio, as noted in previous studies~\cite{Khaitovich:2005gr,Gilad:2006kta,Fraser:2011kt}; however, it does not generate the linear increase $\Omega_\ad (\tau) \simeq \vv \tau/2$ characteristic of fitness peak displacements (Fig.~\ref{fig:demographic}).

\paragraph{Gene-specific relaxation of stabilizing selection.}
We can also test for lineage- and gene-specific relaxation of stabilizing selection on gene expression, which arises, for example, from a partial loss of gene function. We model the loss dynamics by a stochastic process: with a small rate $\gamma$, individual genes switch the stabilizing strength of their fitness seascape to a reduced value $\lambda c$ (with $\lambda < 1$). This dynamics increases cross-species divergence and generates a nonlinear time-dependent $\Omega (\tau)$,  not observed in the data (Fig.~1A). To discriminate between relaxed stabilizing selection and directional selection, we also use the distributions of clade-specific expression differences $\Delta E^\alpha_\C$, which are defined as averages over pairwise differences $\Delta E^\alpha_{ij} = E^\alpha_i - E^\alpha_j$ in analogy to equation (\ref{OmegaC}). The observed distributions  are of approximately Gaussian form, 
\EQ
P_\C (\Delta E) = \frac{1}{\sqrt{2 \pi D_\C}} \exp \left [ - \frac{(\Delta E)^2}{2 D_\C} \right ] \label{Delta_E_dist},
\EE
as shown by the collapse plot of Fig.~\ref{fig:Alternative2}A. This is in accordance with the minimal seascape model, which predicts a Gaussian distribution $P_\tau (\Delta E)$ with variance $\langle D(\tau)\rangle$. In contrast, stochastic relaxation of stabilizing selection generates broad non-Gaussian tails increasing with divergence time $\tau$ that are not observed in the data (Fig.~\ref{fig:Alternative2}C). We conclude that relaxed stabilized selection alone cannot the observed statistics of {\em Drosophila} gene expression levels. This does not exclude that relaxation of selection affects some genes in our  data set and more broadly genes with complete loss of function, which are suppressed in the set of conserved orthologs.

\paragraph{Punctuated directional selection.} 
The Ornstein-Uhlenbeck dynamics of fitness peaks in the minimal seascape model  (equation~\ref{ou}) describes the accumulation of  small but continual changes of optimal expression levels. Larger peak shifts can be caused by discrete ecological events, including major migrations and speciations, and by gene-specific factors such as neo-functionalization~\cite{Lynch:Genetics}. Here we model such events by a {\em punctuated} fitness seascape~\cite{Held:2014ve}: with a small rate $\vv \mu / (2r^2)$, individual genes are subject to fitness peak shifts by an amount of order $E_0$. This stochastic model differs from previous models of lineage-specific selection~\cite{Hansen:1997ws,Butler:2004th,Hansen:2008gt,Bedford:2009fy,Kalinka:2010dw,Brawand:2011,Rohlfs:2014bl}, where fitness peak shifts are constrained to known branch points of the phylogeny. Evolution in a punctuated fitness seascape generates divergence-diversity ratio $\Omega (\tau)$ of the form (equation~\ref{Omega_ad}); adaptation is signalled by the same term $\Omega_\ad (\tau) \simeq \vv \tau/2$ as in a minimal seascape of the same driving rate $\vv$. To discriminate between the two models, we use again the distributions $P_\C (\Delta E)$ of clade-specific expression differences. In a punctuated seascape, these distributions have broad non-Gaussian tails increasing with divergence time $\tau_C$ that are not observed in the data (Fig.~\ref{fig:Alternative2}D). We conclude that large peak shifts are a subleading factor of expression changes in our data set.

\paragraph{Other modes of adaptation.} 
Further evolutionary modes affecting gene expression include: 
\begin{enumerate}[(a)]

\item Time-dependent stabilizing selection~\cite{Held:2014ve}. This type of selection can be modeled by a fitness seascape of the form (\ref{fitness}) with time-dependent stabilizing strength $c(t)$, given by a generalized Ornstein-Uhlenbeck process with constraint $c (t) > c_{\min}$. The recurrent tightening of expression constraint driven by increases of $c(t)$ is a mode of adaptation that is independent of fitness peak changes. The $\Omega$ test does not trace this mode: as long as the expression optimum $E^*$ is time-independent, the function $\Omega (\tau)$ reaches an asymptotic value $ < \Omega_\stab (c_{\min})$. This pattern is similar to evolutionary equilibrium in a single-peak fitness landscape and does not contain the term  $\Omega_\ad (\tau) \simeq \vv \tau/2$ characteristic of fitness peak displacements. 

\item Adaptive gene turnover, including sub- and neo-functionalization after gene duplica\-tion~\cite{Lynch:sub_1,Lynch:sub_2}, regulatory sequence duplication~\cite{Nourmohammad:2011},  and de novo formation of genes~\cite{Tautz:orphan}. This mode is suppressed in our data set of conserved orthologous genes, but it is likely to be more prevalent in the complementary set of {\em Drosophila} genes. 
\end{enumerate}
A detailed investigation of these evolutionary modes is beyond the scope of this study. Importantly, however, they do not confound the inference of adaptation under directional selection reported here.

\subsubsection*{4. Analysis of specific gene classes}

\paragraph{Codon usage.} 
The effective number of codons, $n$, measures the redundancy of the genetic code within a given gene~\cite{Wright:1990tf}. This number takes values between 20 (each amino acid is determined by a specific codon) and 61 (all sense codons are used). Genes with specific codon usage (small $n$) tend to have higher expression than genes with broad codon usage~\cite{Ikemura:1985ui,SHIELDS:1988uj}. Here we compute the species-averaged effective number of codons, $n^\alpha = \frac{1}{7} \sum_i n^\alpha_i$ for all genes in our data set.
 We find a consistent dependence of expression adaptation on codon usage: 
\begin{enumerate}[(a)] 
\item Aggregate analysis by the $\Omega$ test signals strongly reduced adaptation for genes with specific codon usage ($n < 42$) and   an enhanced adaptation for genes with broad codon usage ($n >50$), compared to the average over all  genes (Fig.~2A and Table~\ref{table:fraction}). 

\item The $\Omega$ test also signals strongly reduced adaptation for genes with high average expression level, $ \bar E^\alpha = \frac{1}{7}\sum_i E_i^\alpha >0.9$ (Table~\ref{table:fraction}). Additionally, we compare the fitness flux of a gene to its codon adaptation index (CAI), which  measures the similarity between the codon usage in a specific gene and the codon preference of highly expressed genes~\cite{Sharp:1987}. Consistently, we find a reduced amount of fitness flux in genes with high  codon adaptation index ($\text{CAI}\gtrsim 0.65$); these genes  are likely to be highly expressed.

\item At the level of individual genes, there is a clear correlation between fitness flux $\Phi^\alpha$ and effective codon number $n^\alpha$ (Fig.~2B). 
\end{enumerate}

\paragraph{Inference of adaptive sequence evolution.} 
For the genes in our data set, we estimate the fractions of synonymous and non-synonymous polymorphic nucleotides  ($P_s$ and $P_n$)  from the database of {\em Drosophila melanogaster } Genetic Reference Panel (DGRP)~\cite{Mackay:2012fd}. The corresponding nucleotide divergence measures ($D_s$ and $D_n$) are obtained from sequence alignments between the  {\em D. melanogaster}  and   {\em D. simulans} reference genomes~\cite{DrosophilaGenomesConsortium:2007gf}. The McDonald-Kreitman test~\cite{Kreitman:1991vh,Smith:2002cm} signals adaptive evolution of amino acids if $\alpha_{\rm seq} = (D_n P_s / D_s P_n) - 1 > 0$. Fig.~\ref{fig:MK} shows the distribution of $\alpha_{\rm seq}$ values for classes of genes with different amount of expression adaptation, measured by the fitness flux $\Phi$ (equation \ref{Phi_alpha}). We find no correlation between these statistics. In each class, about $30\%$ of the genes have $\alpha_{\rm seq} > 0$. This result does not contradict the correlation of gene expression divergence and amino acid divergence reported in ref.~\cite{Zhang:2007bf}, because an enhanced amino acid substitution rate measured by a $D_n/D_s$ test~\cite{Li93} may be caused by adaptive changes or by relaxation of negative selection.

\paragraph{Analysis of functional gene classes.}
We use The Ontologizer~\cite{Bauer:2008fz} for statistical analysis of functional enrichment in our dataset. From a base set of all 6332 genes in our database, we identify enriched functional categories in the query sets of adaptively regulated genes ($2N\Phi^\alpha>4$) and genes with sex-specific adaption of expression ($2N \Phi_{\mf}^\alpha>4.5$, see below). We use the calculation method Parent-Child-Union with  Bonferroni correction and  resampling steps of 1000. The enriched functional categories in these gene sets are reported in  Tables S1 and S2 with a significance threshold  $P<0.1$ (multiple hypothesis test). We list three main  categories: biological processes, cellular components, and molecular functions. Each functional category is assigned to a functional cluster (in bold letters) that is inferred by REVIGO~\cite{Supek:2011bw}, using the semantic similarity measure SimRel with threshold 0.5. This clustering facilitates the interpretation of functional gene classes associated with adaptation of gene expression.

\paragraph{Sex-specific evolution and sex bias of expression.}  
To quantify  differences of gene expression between male and female individuals, we we define the sex specificity trait of a given gene as the difference between its expression levels in males and in females~\cite{Zhang:2007bf}, 
\EQ
E^\alpha_{\mf, i} = E^\alpha_{m,i} - E^\alpha_{f,i}. 
\EE
We analyze these traits by the same methods as the sex-averaged expression levels $E_i^\alpha$ defined by equation (\ref{Eia}). Specifically, we define the aggregate divergence-diversity ratio $\Omega_{\mf, \C}$ and the fitness flux $2N \Phi_\mf$ in analogy to equations (\ref{OmegaC}) and (\ref{Phidef}), and we infer gene-specific maximum-likelihood values $2N \Phi_\mf^\alpha$ in analogy  to equation (\ref{fitness}). We define two conceptually distinct measures of male-female differentiation:

\begin{enumerate}[(a)]

\item Sex-specific adaptation. In accordance with ref.~\cite{Zhang:2007bf}, we find that most genes of our data set have well-conserved and often small sex specificity; these genes evolve their expression levels coherently between males and females. We use the rescaled fitness flux $2N \Phi_\mf$ to delineate coherent evolution of expression levels (i.e., conservation of the specificity trait) from sex-specific adaptation (i.e., adaptive changes of the male-female expression difference), as illustrated in Fig.~2C. A set of 1155 sex-specific adaptive genes is identified by the condition $2N \Phi_\mf^\alpha > 4.5$ (Table~S2); we use a more stringent threshold than for $\Phi^\alpha$ because the sex-specificity trait statistics has larger statistical errors.

\item Sex bias. We identify genes with male- and female-biased expression in {\em Drosophila} using the results of Assis et al.~\cite{Assis:2012ez}, which are based on a number of statistical tests in the whole body and in gonads of males and females in {\em D. melanogaster} and {\em D. pseudoobscura}. A gene is classified as expression sex-biased if flagged by at least three of these tests, which  produces a list of 450 male-biased and 499 female-biased  genes. A related measure of bias within our data set is the 
species-averaged specificity trait, $\bar E_\mf^\alpha = \frac{1}{7} \sum_i E^\alpha_{\mf, i}$.  
\end{enumerate}
Our analysis establishes a relation between these two measures in our data set: strong sex-specific adaptation of expression occurs in male-biased, but not in female-biased genes. First, the aggregate ratio $\Omega_{\mf}$ in male-biased genes show evidence for adaptive evolution with a linear adaptive component $\vv\tau/2$. Unbiased and female-biased genes have only a small average divergence in their sex-specificity trait that is of the order of the expression diversity (i.e., they within the error range of the observed expression levels), providing no evidence for adaptation (Fig.~2D). Second, the fitness flux $\Phi^\alpha_\mf$ is strongly enhanced for genes with large $\bar E^\alpha_\mf$ (Fig.~2E). Accordingly, 32\% of male-biased genes are classified as sex-specific adaptive. Functional categories associated with sex-specific adaptation of expression are reported in Table~S2.

\subsubsection*{5. Simulation tests}
\paragraph{In-silico evolution of quantitative traits.} 
We use a  Fisher-Wright process  for the evolution for the evolution of populations along the Drosophila phylogeny of Fig.~\ref{clade_phyl}A. A population consists of $N$ individuals with genomes $\a^{(1)}, \dots, \a^{(N)}$. A genotype is an $\ell$-letter sequence $\a = (a_1, \dots, a_\ell)$ with alleles $a_\i = 0,1$ ($\i = 1, \dots, \ell)$. It defines an expression level $E(\a) = \sum_{\i =1}^\ell \cE_\i a_\i$ with neutral variance $E_0^2 = \frac{1}{4} \sum_{\i=1}^\ell \cE_\i^2$. We use uniform single-locus effects $\cE_i$; our results are insensitive to the form of the effect distribution~\cite{Held:2014ve}. In each generation, the sequences undergo point mutations with a probability $ \mu \tau_0$ per generation, where $\tau_0$ is the generation time. The sequences of next generation are then obtained by multinomial sampling with a probability proportional to $[1 + \tau_0 f(E(\a),t)]$, where the fitness function $f(E, t)$ is given by equation (\ref{fitness}). Simulations are performed with $N=100$, $\pi_0=0.1$ for traits with $\ell = 100$, uniform effects $E_i = 1$, and average fitness optimum $\mathcal{E}=70$. We use three different types of selection (for details, see ref.~\cite{Held:2014ve}):
\begin{enumerate}[(a)]
\item Minimal fitness seascape. Before each reproduction step, a new optimal trait value $E^*(t + \tau_0)$ is drawn  from a Gaussian distribution with mean $E^*(t)(1- \mu \tau_0 \vv / (2r^2))+ \mathcal E \, \mu \tau_0 \vv / (2r^2)$ and variance $\mu \tau_0 \vv E_0^2$. 

\item Fitness landscape. The optimal trait value $E^*$ is time-independent (Fig.~\ref{fig:demographic}A,B). In the model of gene-specific relaxation of selection (see section~3), the stabilizing strength of individual genes switches to a smaller value, $c \to 0.01 c$, with a small rate $\gamma$ (Fig.~\ref{fig:Alternative2}C). 

\item Punctuated fitness seascape (see section 3). Before each reproduction step, a new, uncorrelated optimal trait value is drawn with probability $\mu \tau_0 \vv /(2r^2)$ from a Gaussian distribution with variance $r^2 E_0^2$, where $r^2$ is a constant of order 1 (Fig.~\ref{fig:Alternative2}D). 
\end{enumerate}

\paragraph{Validation of the probabilistic inference scheme.}
To test the performance of our inference scheme, we generate expression values $\E^\alpha = (E_1^\alpha , \dots, E_7^\alpha)$ for individual genes with trait scales $E_{0,\alpha}^2$ by Fisher-Wright simulations along the {\em Drosophila} phylogeny of Fig.~\ref{clade_phyl}A. We use minimal fitness seascapes of the form (\ref{fitness}) with input parameters $(c_{\rm in}, \vv_{\rm in})$ and a sequence diversity $\pi_0 = 4 \mu N = 0.05$. We then infer maximum-likelihood posterior values $(c^\alpha, 2N \Phi^\alpha)$ by the probabilistic method described in section~2 (equation~\ref{Phi_alpha}). In Fig.~\ref{fig:sim}A, we plot the distribution of inferred fitness flux values $2N \Phi^\alpha$ against the input expectation value $2N \Phi_{\rm in} = 2 c_{\rm in} \vv_{\rm in} \, \tau_{\rm Dros.}$ (equation~\ref{Phicv}). The underlying simulations use a range of trait scales $E_{0,\alpha}^2 = 0.25-4.0$ appropriate for log expression levels; the inference of $\Phi^\alpha$ does not require knowledge of this scale (see section~2). Fig.~\ref{fig:sim}B shows the corresponding distribution of inferred values $c^\alpha$ as a function of the input stabilizing strength $c_{\rm in}$. These simulations use a uniform trait scale $E_{0,\alpha}^2 = 1$ (inferring the actual scales requires sufficiently reliable gene-specific expression diversity data). The posterior values $(\Phi^\alpha, c^\alpha)$ are seen to provide reasonable, on average conservative estimates of the input model parameters $(c_{\rm in}, \Phi_{\rm in})$. In particular, the inference of a significant fitness flux ($\Phi > 1/2N)$ is incompatible with evolution under static stabilizing selection ($ \vv=0,\, c>0$) or near neutrality ($c \simeq 1$), independently of the underlying model for the adaptive evolution of a  molecular trait.

\paragraph{Robustness of the inference to trait epistasis.} 
The analytical theory underlying our inference method~\cite{Nourmohammad:2013ty,Held:2014ve} covers molecular quantitative traits with a linear genotype-phenotype map, $E(\a) = \sum_{\i =1}^\ell \cE_\i a_\i$ (see above). Here we extend this method to nonlinear traits of the form $E(\a) = \sum_{\i=1}^\ell \cE_\i a_\i + \sum_{\i < \i'} \cE_{\i \i'} a_\i a_{\i'}$; such nonlinearities are commonly referred to as trait epistasis. The strength of epistasis can be defined as the ratio of nonlinear and linear neutral trait variance, $\varepsilon^2 =  (\sum_{\i < \i'} \cE_{\i \i'}^2) / (\sum_\i \cE_\i^2)$. 

Trait epistasis introduces only minor changes to the quantitative genetics theory of refs.~\cite{Nourmohammad:2013ty,Held:2014ve}. In particular, the $\Omega$ ratio retains its normalization in the neutral long-term limit and the quasi-neutral growth of the trait divergence is still given by equation (\ref{qn}), where $\Delta$ is now the total genetic diversity of the trait. To specifically test our inference method, we perform Fisher-Wright simulations as described above over a wide range of the parameter $\varepsilon^2$; individual  epistatic effects $\cE_{\i \i'}$ are drawn from a Gaussian distribution with mean 0. In an ensemble of 6000 independently evolving traits, we record both the actual average fitness flux (equation~\ref{Phidef}) and the inferred fitness flux determined from the aggregate $\Omega$ ratio (equation~\ref{omegaPhi}). Both quantities show no systematic dependence on $\varepsilon^2$ (Fig.~\ref{fig:sim}C), suggesting that our inference of adaptive evolution is not confounded by trait epistasis.

{\small

\paragraph{\small Acknowledgments.}
We acknowledge discussions with P. Andolfatto, N. Barton, M. \L{}uksza, V. Mustonen, P. Shah, and S. Schiffels.
This work has been partially supported by James S. McDonnell Foundation $21^{st}$ century science initiative-postdoctoral program in complexity science / complex systems (AN) and and by Deutsche Forschungsgemeinschaft grant SFB 680. This work has also been supported by the Gordon and Betty Moore Foundation grant no. 2919 and National Institutes of Health grant no. R25 GM067110 to the Kavli Institute of Theoretical Physics (Santa Barbara), where part of this work was performed.

}

\newpage

\setcounter{figure}{0}
        \renewcommand{\thefigure}{S\arabic{figure}}
        \renewcommand{\thetable}{\arabic{table}}
               
\vspace*{2cm}
\begin{figure}[h]
\begin{center}
\includegraphics[width= \textwidth]{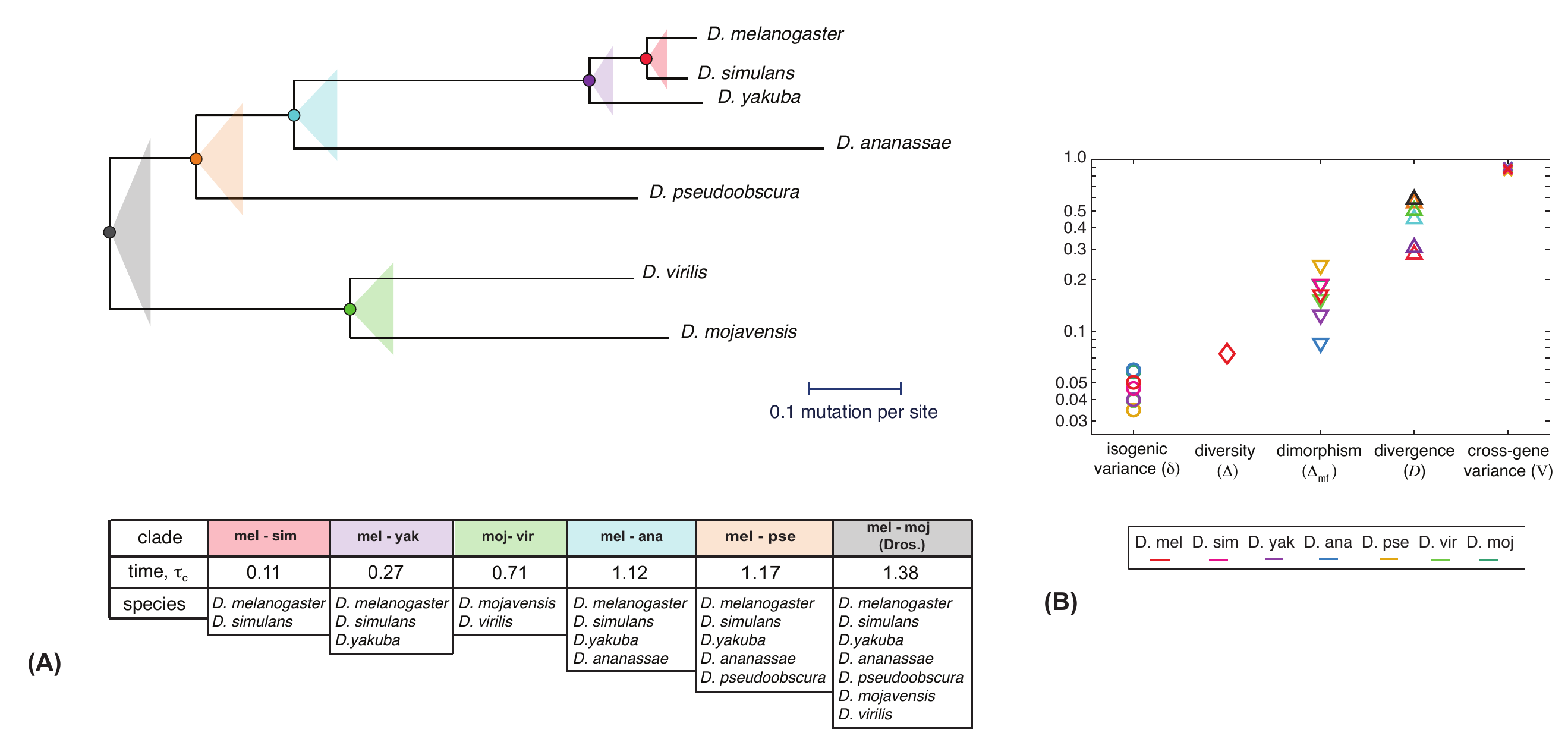}
\end{center} 
\caption{{\bf  Species phylogeny and levels of gene expression variation.} {\bf (A)} Phylogeny of the Drosophila genus, as reconstructed in ref.~\cite{DrosophilaGenomesConsortium:2007gf} from synonymous sequence divergence. Six clades are marked by 
colored triangles, their ancestral nodes by colored circles. The table specifies the species contained each of the clades and the clade divergence time $\tau_\C$ (equation~\ref{tauC}). {\bf (B)} Gene-averaged isogenic variance $\langle\delta\rangle$ ($\circ$, equation~\ref{isogenic_statistics}), expression diversity $\langle\Delta\rangle$ ($\bigtriangledown$, equation~\ref{Delta}), male-female expression dimorphism $\langle \Delta_{mf}\rangle$ ($\diamond$, equation~\ref{Delta_mf}), clade divergence $\langle D\rangle$ ($\triangle$, equation~\ref{D}),  and cross-gene variance of expression $V=\langle \Gamma_i^2 \rangle\approx 1$ ($\times$).  We find a clear ranking $\langle \delta_i \rangle < \langle \Delta \rangle\lesssim\langle\Delta_{mf}\rangle < \langle D_{ij} \rangle < V_i$. The color code for single-species data is shown in the legend, colors for clades are as in (A).}
\label{clade_phyl} 
\end{figure}


\newpage{}

\begin{figure}[c]
\begin{center}
\includegraphics[width= \textwidth]{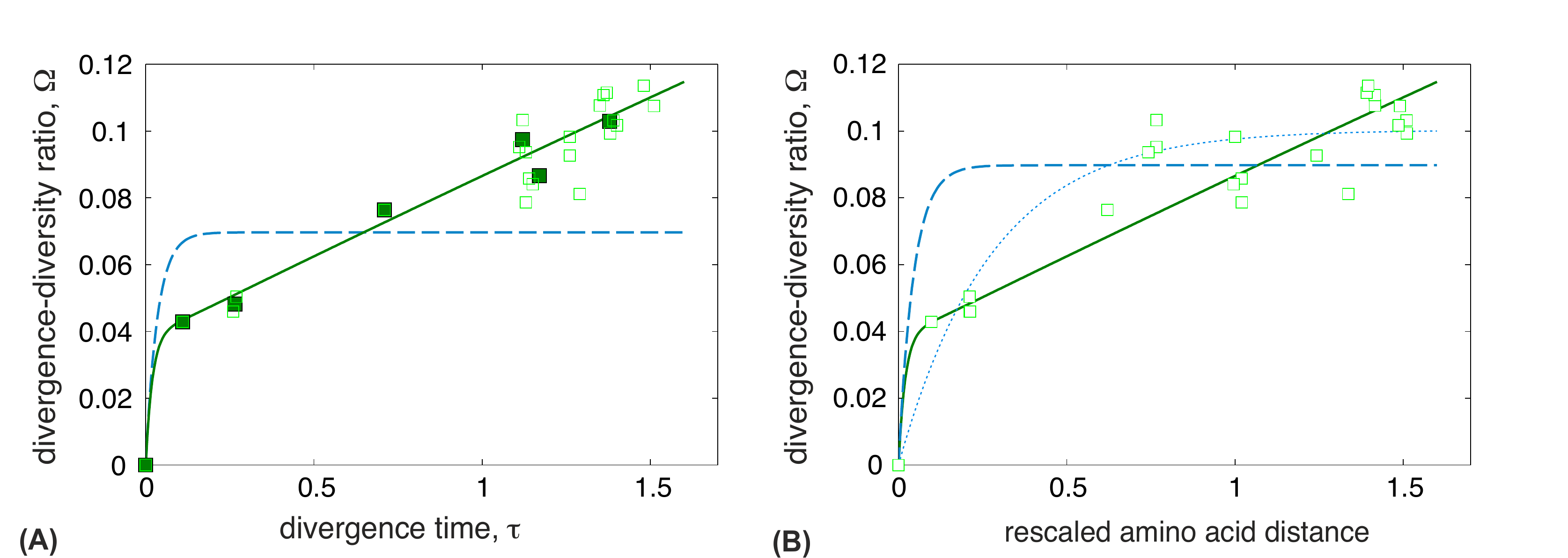}
\end{center} 
\caption{{\bf Fitness landscape and Ornstein-Uhlenbeck models as control.} {\bf (A)} The aggregate divergence-diversity ratio for clades, $\Omega_\C$ (filled squares), is compared to the ratio for individual pairs of species, $\Omega_{ij}$(empty squares). Both quantities are plotted against the divergence time estimated from four-fold synonymous sites~\cite{DrosophilaGenomesConsortium:2007gf} (Fig.~\ref{clade_phyl}A). Clade-based statistics is seen to substantially reduce the noise of the expression divergence data. The seascape model (green line as in Fig.~1A; $c^* = 18.6$, $\vv^* = 0.07$) provides a significantly better fit to these data than the landscape model (blue line; $c_\eq = 13$); see Fig.~1C and section~2 of Materials and Methods for a statistical model comparison.  
{\bf (B)} The same species-pair ratio $\Omega_{ij}$ (empty squares) is plotted against the amino-acid sequence distance used in ref.~\cite{Bedford:2009fy}, uniformly rescaled to give the same scaled genus divergence time $\tau_{\text{Dros.}}=1.4$ as in (A). Compared to synonymous divergence, amino-sequence divergence is seen to produce an inhomogeneous molecular clock that adds to the noise of the evolutionary analysis. For these data, the seascape model (green line as in (A)), the landscape model (dashed blue line; $c = 6.4$), and the Ornstein-Uhlenbeck model~\cite{Hansen:1997ws} (blue line; $ \lambda = 1.8 \,\mu^{-1} $, $\sigma^2 = 0.13 \, \langle\Delta\rangle/N$; see equation~\ref{ou} and ref.~\cite{Bedford:2009fy}) produce fits of overall comparable quality. However, both equilibrium models cannot explain the evolution of expression in the youngest clades: the landscape model overestimates the divergence $D_{ {mel}-{yak}}$ and $D_{ {mel}-{sim}}$, the  Ornstein-Uhlenbeck model overestimates the relative divergence $D_{ {mel}-{yak}} / D_{ {mel}-{sim}}$. See section~2 of  Materials and Methods. 
\label{fig:Hartl}
}
\end{figure}


\newpage
\begin{figure}[c]
\begin{center}
\includegraphics[width= 1 \textwidth]{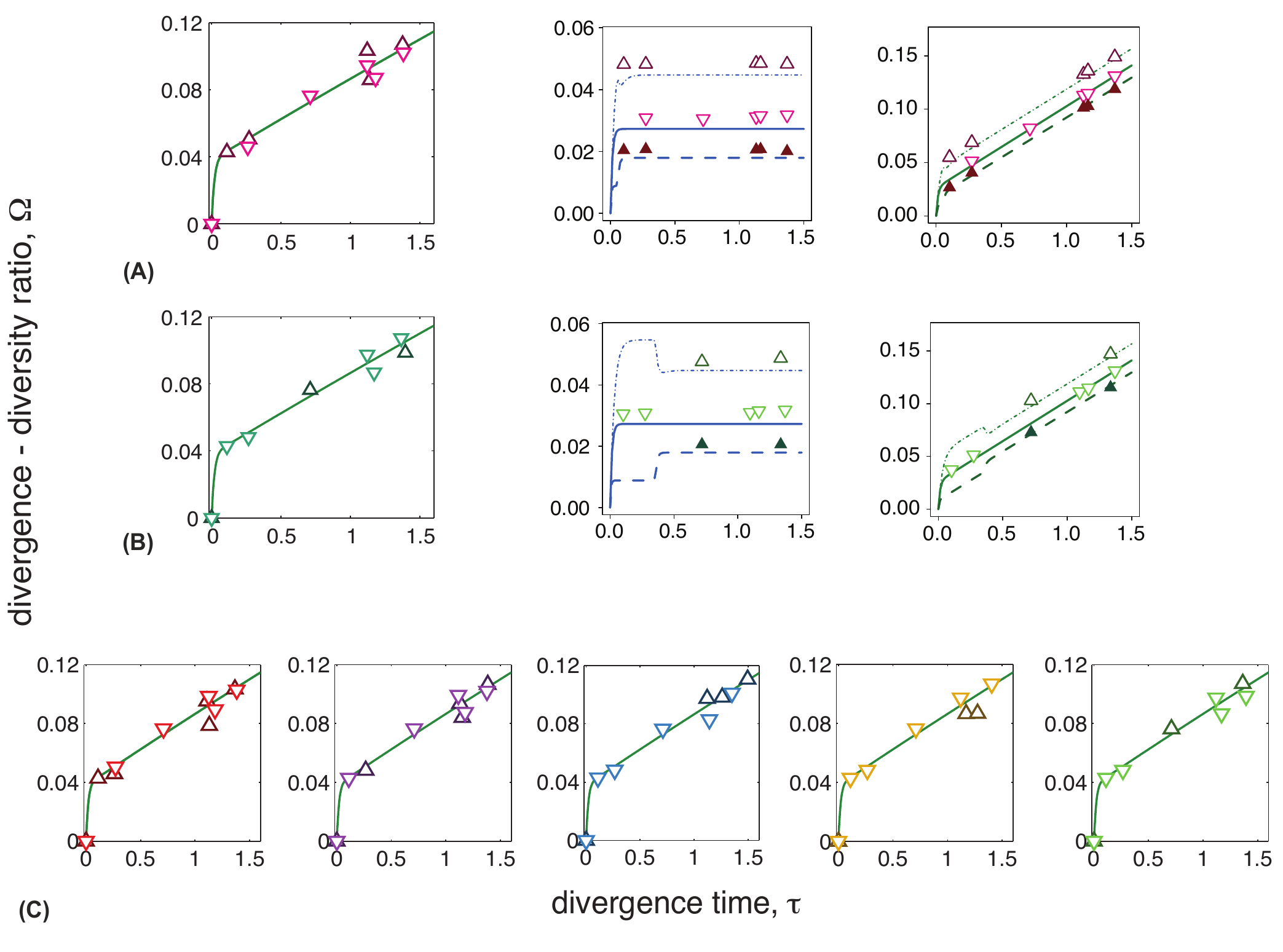} 
\end{center} 
\caption{{\bf Test of lineage-specific demography.} We compare the polarized divergence-diversity ratio $\Omega_{\C,i}$ with species $i$ as outgroup (equation \ref{OmegaCi}, $\triangle$) to background data from partial clades {\em excluding} species $i$ ($\bigtriangledown$); both quantities are plotted against the clade divergence time. 
{\bf (A)}~Left panel: Data for clades with outgroup \emph{D. melanogaster}. Center and right panels: Evolution with a reduced or enhanced effective population size $N_i$ in the outgroup lineage. Analytical curves and simulation results are shown for  $N_i=3N$ (dashed lines, $\blacktriangle$) $N_i = N/2$ (dashed-dotted lines, $\triangle$) in a fitness landscape ($c = 20, \vv = 0$; center panel) and seascape ($c = 20, \vv = 0.09$; right panel). 
{\bf (B)} Same as (A), with outgroup {\em D. mojavensis}. 
{\bf (C)} Data for each of the other five species chosen as outgroup. These data give no evidence of long-term lineage-specific demography. The analytical and simulation results show that lineage-specific demography under stabilizing selection does not give a spurious signal of adaptive evolution in the $\Omega$ test. Lineage-specific demography is introduced in section 3, simulation details are given in section~5 of  Materials and Methods.}
 \label{fig:demographic}
\end{figure}

\newpage
\begin{figure}[c]
\begin{center}
\includegraphics[width=\textwidth]{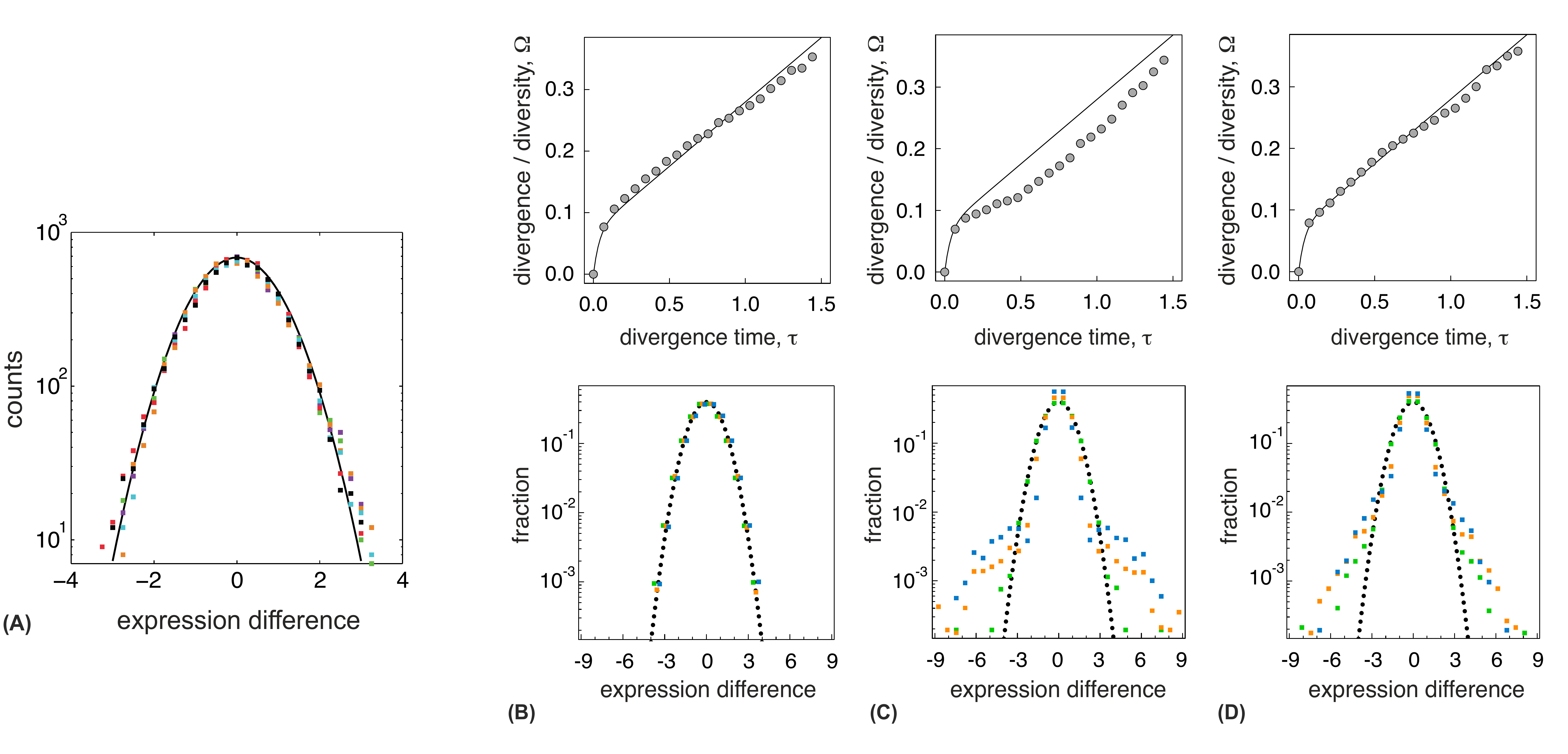}
\end{center} 
\caption{{\bf Test of alternative selection scenarios.} 
{\bf (A)} Distributions of clade-specific expression level differences,  $P_\C (\Delta E)$ (equation~\ref{Delta_E_dist}, color code as in Fig.~\ref{clade_phyl}A), standard-normalized to mean 0 and variance 1. These distributions are approximately Gaussian (black line: standard normal distribution).   
{\bf (B)} Minimal seascape model. Top panel: Divergence-diversity ratio $\Omega(\tau)$ (bullets: simulation results; line: analytical curve as in Fig.~1A). Bottom panel:  Standard-normalized  distributions of trait differences, $P_\tau (\Delta E)$, from simulations for $\tau = 0.21, 0.69$ and $1.37$ (green, orange, and blue bullets) are of Gaussian form (dotted line). The same quantities are shown for two alternative fitness models: 
{\bf (C)} Loss-of-function model. Functional genes evolve in a static fitness landscape of stabilizing strength $c = 4.5$; individual genes lose function with rate  $\gamma = 0.04 \,\mu$, resulting in reduced selection ($c \rightarrow 0.01 \,c$). The loss events generate a nonlinearity in $\Omega(\tau)$ and a broad tail in  $P_\tau (\Delta E)$ that are not observed in the data. 
{\bf (D)} Punctuated fitness seascape. Individual genes jump to a new, uncorrelated fitness peak with rate $0.16 \, \mu$. These dynamics also generate a broad tail in  $P_\tau (\Delta E)$. The {\em Drosophila} data of $\Omega_\C$ (Fig.~1A) and of $P_\C (\Delta E)$ together favor the minimal seascape model over both alternatives. The loss-of-function model and the punctuated seascape model are introduced in section~3, simulation details are given in section~5 of  Materials and Methods. }
\label{fig:Alternative2}
\end{figure}

\newpage
\begin{figure}[c]
\begin{center}
\includegraphics[width=  \textwidth]{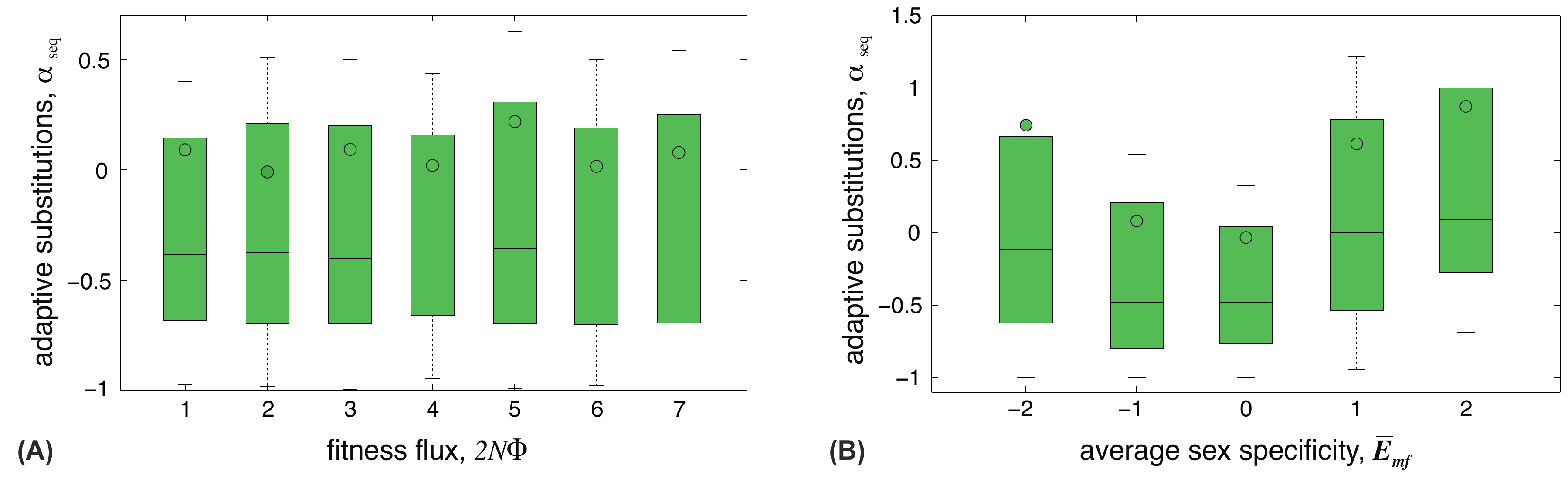}
\end{center} 
\caption{{\bf Adaptive gene expression versus adaptive evolution of protein sequence.} 
(a) The distribution of  $ \alpha_{\rm seq} = (D_n P_s / D_s P_n) - 1$, denoting the fraction of adaptive amino acid substitutions~\cite{Smith:2002cm},  is plotted against the cumulative fitness flux of gene expression, $2N\Phi$ (circle: average; line: median; box:  $50\%$ around median; bars: $70\%$ around median). We find no correlation  between these statistics, which suggests that adaptive gene expression is an independent mode of evolution. 
(B) The distribution of $ \alpha_{\rm seq}$ plotted against the average  sex specificity  $\bar E_\mf$ signals increased adaptive protein evolution in genes with sex-biased expression, which is strongest in male-biased genes (cf.~the results of ref.~\cite{Zhang:2007bf}). For the definition of sex-biased expression, see section~4 of Materials and Methods.} 
 \label{fig:MK}  
\end{figure}
\clearpage

\newpage
\begin{figure}[c]
\begin{center}
\includegraphics[width=  \textwidth]{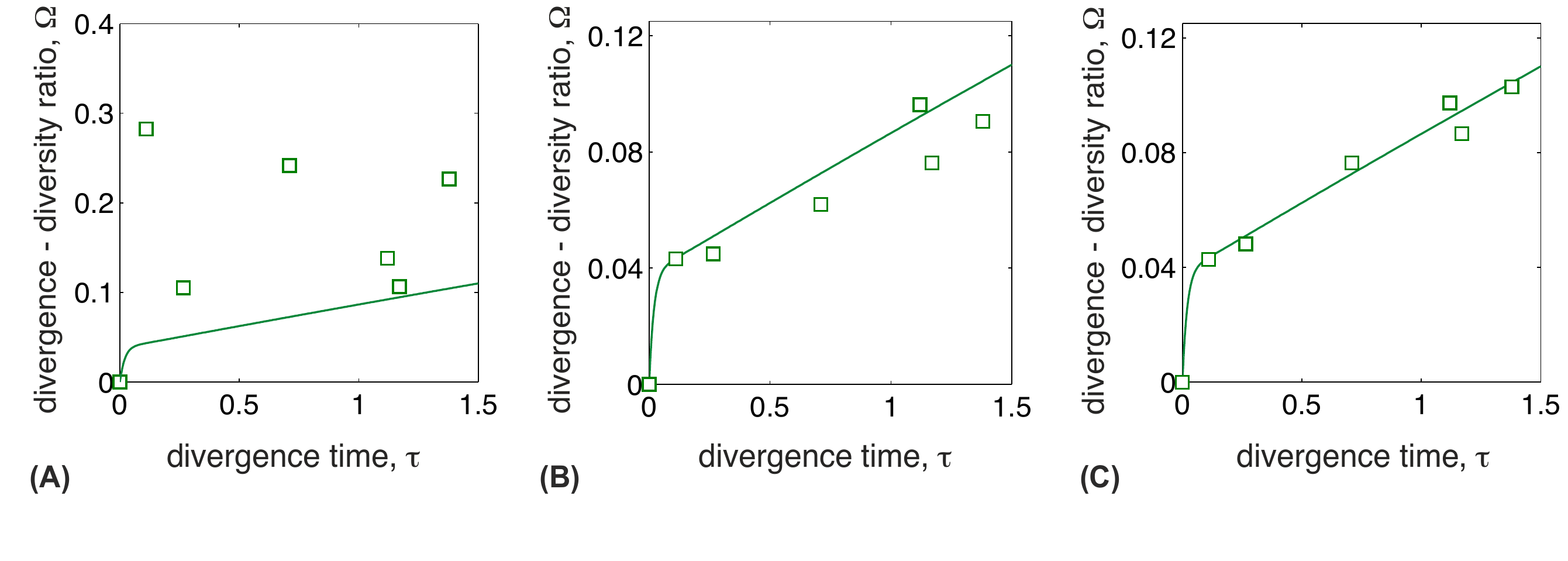}\hfill{} \\
\end{center}
\caption{{\bf Rescaling of gene expression data.}  
The clade-specific divergence-diversity statistics $(\Omega_\C, \tau_\C)$ are shown for three variants of expression levels. 
{\bf (A)} Raw data: $\log_2$ intensities averaged over experimental replicates. 
{\bf (B)} Levels shifted to gene average 0 within each replicate line. 
{\bf (C)} Levels shifted to gene average 0 and normalized to variance 1 within each  replicate line~\cite{Quackenbush:2002kl}, as used throughout this paper. The theoretical curve $\Omega (\tau)$ of the best-fit seascape model ($c^* = 18.6, 2N \Phi^* = 3.6$) is shown in all three panels. The additive shift in (B,C) is seen to be essential for evolutionary analysis, the subsequent multiplicative rescaling in (C) further reduces the noise in the $\Omega$ data. See section~1 of  Materials and Methods.
\label{fig:rescaling}}
\end{figure}
\clearpage

\newpage
\begin{figure}[c]
\begin{center}
\includegraphics[width= \textwidth]{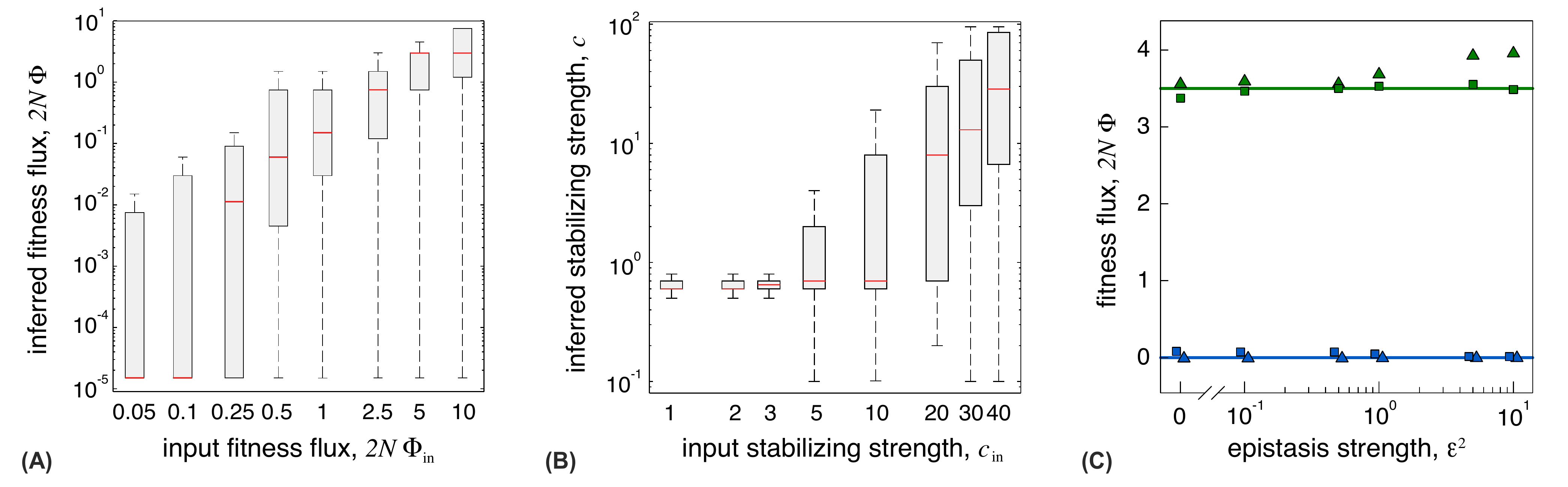}
\end{center} 
\caption{{\bf Simulation tests of the inference scheme.} 
{\bf (A,B)} Distributions of the cumulative fitness flux  $2N\Phi^\alpha$ and stabilizing strength $c^\alpha$ inferred from simulated expression data are plotted against the simulation input parameters $2N \Phi_{\rm in}$ and $c_{\rm in}$ (red line: median, box: $50\%$ around the median, bar: $75\%$ around median). See section~5 of  Materials and Methods for simulation details. 
{\bf (C)} Selection inference for epistatic traits. Simulation results of the actual fitness flux ($\triangle$) are compared to  flux values inferred by the standard $\Omega$ test~($\square$, see section~2 of Methods). Both quantities are plotted against the strength of epistasis, $\varepsilon^2$, defined as the ratio of epistatic and additive trait variance (section~5 of Methods); horizontal lines show the actual fitness flux without epistasis ($\varepsilon^2=0$). Simulations are shown for selection parameters $(c = 4.5, \vv = 0.4)$ (green) and $(c = 4.5, \vv = 0)$ (blue). See section~5 of  Materials and Methods for simulation details. 
\label{fig:sim}}
\end{figure}

\clearpage

\setcounter{table}{0}

\newpage
{\small
\begin{table}[c]
\begin{footnotesize}
\begin{center}
\begin{tabular}{ l || c | c | c | c | c | c | c | c | c }
\multicolumn{3}{c}{} & \scriptsize{mel-sim} & \scriptsize{mel-yak} & \scriptsize{vir-moj} & \scriptsize{mel-ana} & \scriptsize{mel-pse} &  \multicolumn{2}{c}{\scriptsize{mel-moj ({\bf Dros.})} }\\ \hline 
{gene class  (gene number) }    &  $c^*$ & $ 2N \Phi^*$ & \multicolumn{5}{c}{$\omega$} & & $\alpha$  
\\ \hline \hline
{\bf all genes} (6332)  & 18.6 & 3.6 & 8\%& 23\%& 47\%& 59\%& 60\%& 63\%&  54\%
\\ \hline
{broad codon usage}   (1176) & 15.6 & 3.9  & 9\%& 25\%& 49\%& 61\%& 62\%& 66\%  & 57\%
\\ \hline
{narrow codon usage}  (501) & 19.0 & 2.3& 5\%& 17\%& 36\%& 48\%& 49\%& 53\%&  18\%
\\ \hline
{high expression} (553)& 14.3 & 1.7 & 1\%& 8\%& 27\%& 39\%& 40\%& 44\%  & 0\%
\\ \hline \hline
\end{tabular}
\end{center}
\end{footnotesize}
\vspace{0.5cm}
\caption{{\bf Selection parameters and amount of adaptation.}
$c^*$: maximum-likelihood stabilizing strength.
$2N \Phi^*$: maximum-likelihood fitness flux.
$\omega$: clade-dependent adaptive fraction of the gene expression divergence.
$\alpha$: fraction of adaptively regulated genes across the {\em Drosophila} genus, given by the condition $2N\Phi^\alpha>4$. See equations (\ref{Phi_alpha}) and (\ref{ML_c_phi}) for definitions; gene classes are introduced in section~4 of  Materials and Methods.
\label{table:fraction}
}
\end{table}
}

\end{document}